\documentclass[preprint,floatfix,eqsecnum,aps,nofootinbib]{revtex4}

\usepackage{amsmath}
\usepackage{amssymb}
\usepackage{amscd}
\usepackage{mathrsfs}
\usepackage{graphicx}
\usepackage{subfigure}
\usepackage{slashed}
\usepackage{mathbbol}
\usepackage{subfigure}
\usepackage{setspace}
\usepackage{fancyheadings}
\usepackage{multirow}
\usepackage{longtable}
\usepackage{array}

\allowdisplaybreaks[1]

\begin{document}

\title{Origin of neutrino masses at the LHC: $\Delta L = 2$ effective operators and their ultraviolet completions}
\author{Paul W. Angel}\email{pangel@student.unimelb.edu.au}
\affiliation{ARC Centre of Excellence for Particle Physics at the Terascale, School of Physics, The University of Melbourne, Victoria 3010, Australia}
\author{Nicholas L. Rodd}\email{nrodd@student.unimelb.edu.au}
\affiliation{ARC Centre of Excellence for Particle Physics at the Terascale, School of Physics, The University of Melbourne, Victoria 3010, Australia}
\author{Raymond R. Volkas}\email{raymondv@unimelb.edu.au}
\affiliation{ARC Centre of Excellence for Particle Physics at the Terascale, School of Physics, The University of Melbourne, Victoria 3010, Australia}

\begin{abstract}
Neutrino masses and mixings can be generated in many different ways, with some of these scenarios featuring new physics at energy scales relevant for Large Hadron Collider searches.  A systematic approach to constructing a large class of models for Majorana neutrinos may be founded upon a list of gauge-invariant effective operators -- formed from quarks, leptons and the Higgs doublet --  that violate lepton-number conservation by two units.  By opening up these operators in all possible ways consistent with some minimality assumptions, a complete catalogue of a class of minimal radiative neutrino mass models may be produced.  In this paper we present an analysis of Feynman diagram topologies relevant for the ultra-violet completions of these effective operators and collect these into a simple recipe that can be used to generate radiative neutrino mass models.  Since high mass-dimension effective operators are suppressed by powers of the scale of new physics, many of the resulting models can be meaningfully tested at the Large Hadron Collider.
\end{abstract}

\maketitle

%--------------------------------------------------------------------------
%--------------------------------------------------------------------------
% INTRODUCTION
%--------------------------------------------------------------------------
%--------------------------------------------------------------------------
\section{Introduction}
\label{sec:intro}

Empirical evidence for new physics is provided by the discovery of neutrino oscillations, the dark matter problem, and the mystery of the cosmological matter-antimatter asymmetry.  This paper will be concerned with the first of these.  The neutrino oscillation data are nicely consistent with the standard idea that neutrinos are massive and non-degenerate, and that there is a unitary mixing matrix relating the neutrino flavour and mass eigenstates.  The discovery of neutrino oscillations is thus also the discovery of non-degenerate neutrino masses and nonzero mixings.  (For the sake of brevity, in the rest of this paper the term ``neutrino masses" will be taken to include non-zero mixing angles when the context is appropriate.)

Neutrino masses imply new physics, because any mechanism for generating the masses requires degrees of freedom beyond those in the minimal standard model (SM).  Although unattractive, it is technically possible that neutrinos are Dirac particles and gain mass in exactly the same way as the other fermions.  But then at least two right-handed neutrino flavours must be added to the minimal SM.  The minimal ways of generating Majorana neutrinos are the type I and II see-saw mechanisms; the former requires at least two right-handed neutrino flavours, and the latter a Higgs triplet.

The new physics may, unfortunately, be essentially impossible to identify.  For example, the minimal and elegant type I see-saw scenario sees the new degrees of freedom as being extremely massive fermions that are singlets under the SM gauge group \cite{type1}.  The direct discovery of such particles seems unlikely in practice.\footnote{Discovery would require the existence of a suitable new gauge force, such as right-handed weak interactions, at the TeV scale in addition to the heavy neutral fermion masses being at that same scale.  The so-called $\nu$SM, which uses the type I see-saw Lagrangian in a different parameter regime, can be tested by looking for keV-scale sterile neutrinos in hadron decays \cite{nuSM}.}

But there are other, more robustly testable schemes, even though a sacrifice in elegance and minimality must usually be accepted.  The type II \cite{type2} and III \cite{type3} see-saw mechanisms at least have the new particles charged under SM electroweak forces, so provided the new physics mass scale is not above a TeV direct discovery at the Large Hadron Collider (LHC) is possible.  But a TeV scale for this new physics is not favoured, because the see-saw argument naturally leads to a much higher scale being preferred.

Radiative neutrino mass models, where the smallness of neutrino masses is connected with their origin being at loop level, are intrinsically more testable than tree-level schemes such as the three see-saw models.  This class of models will be our focus in this paper.  These models are more testable for a few reasons.  First, the suppression due to the mass scale $M$ of the new physics is stronger than the $1/M$ of the standard see-saw.  Second, there is an automatic $1/16\pi^2$ suppression per loop.  Third, the neutrino self-energy graph will contain the product of a few dimensionless coupling constants, and if each of them is below unity then further suppression results.  Some of these will be the known electroweak Yukawa coupling constants, which are all much less than one except for the top quark case.  Furthermore some Yukawa coupling constants involving exotic scalars and/or fermions may need to be small to satisfy flavour-changing process bounds.

A subset of phenomenologically acceptable radiative neutrino mass models may even be \emph{falsifiable} at the LHC.  We know that at least one neutrino mass eigenvalue must be no smaller than about $0.05$ eV, otherwise the atmospheric and long baseline $\nu_\mu$ disappearance effects cannot be understood \cite{Osc}. If the suppression due to powers of $1/M$ is sufficiently strong, then to meet the $0.05$ eV lower bound the new physics may be required to be no higher in scale than a TeV.  Some models are actually already ruled out, such as 4- or higher-loop models of neutrino mass, as the new physics should already have been discovered.

It scarcely needs saying that falsifiable models of neutrino mass are worth having.  You are either going to get lucky, or you will rule out logical possibilities.  By ruling out logical possibilities, you increase the likelihood that any of the remaining models is correct.  In the end, it may well be that a high-scale type I see-saw mechanism operates in nature, and while we may never be able to prove it, we can gain circumstantial evidence for it by ruling out alternatives.  Those radiative neutrino mass models that are not falsifiable at the LHC will nevertheless be meaningfully constrained.

Many different radiative neutrino mass models seem to be \emph{a priori} possible, only a few of which have been thoroughly analysed in the literature.  The search through this ``theory space'' calls for a systematic approach.  The way forward is revealed by reviewing the origin of the three famous tree-level see-saw models.  Their basis is the unique (up to family combinations) gauge-invariant mass-dimension five operator that can be constructed out of standard model fields: the Weinberg operator,
\begin{equation}
{\cal O}_1 = LLHH,
\end{equation}
where $L$ is the lepton doublet and $H$ is the Higgs doublet \cite{O1}. The $LL$ notation is short for $\overline{(L_L)^c} L_L$ where $L_L = (\nu_L, e_L)^T$, and appropriate SU(2) index contractions are understood.  This effective operator breaks lepton-number conservation by two units, and becomes a Majorana neutrino mass $m_\nu$ of order $v^2/M$ when the Higgs field is replaced with its vacuum expectation value (VEV), $\langle H \rangle = v$. The inverse relationship between $m_\nu$ and the scale of new physics $M$ is the essence of the see-saw effect.  This non-renormalisable operator can be ``opened up"  -- derived from an underlying ultraviolet (UV) complete or renormalisable theory -- in three different ways at tree-level.\footnote{There are three \emph{minimal} ways.  Interesting non-minimal UV completions also exist.}    These three ways correspond exactly with the type I, II and III see-saw models.  By starting with this effective operator and UV completing it, at tree-level, in all possible minimal ways, one arrives at the three logically possible see-saw models. This process can be replicated for higher mass-dimension $\Delta L = 2$ gauge-invariant effective operators.  

One class of such operators is simply given by ${\cal O}_1 (\overline{H} H)^n$, where $n = 1, 2, 3,\ldots$.  These higher-order versions of the Weinberg operator provide a neutrino mass of order $v^{2(1+n)}/M^{1+2n}$.  They are of interest because the enhanced suppression requires $M$ to generically be a lower scale than for ${\cal O}_1$ models, so the underlying UV complete theories are more testable than the standard see-saw models.  Since $\overline{H} H$ is a singlet under any internal symmetry, any model that yields an $n \ge 1$ operator as the dominant one must be constructed to be somehow unable to generate ${\cal O}_1$, even though the latter would be allowed by all the internal symmetries of that model.  One approach is to break minimality by having multiple Higgs doublets $H_i$, such that $\overline{H}_i H_j$ is not an internal symmetry singlet when $i \neq j$.  Another is to invoke supersymmetry. For a systematic treatment of this approach up to the order of 1-loop models see \cite{Winter1,Winter2,KO}.

But we are concerned here with operators that have a structure completely different from ${\cal O}_1$, but maintain the $\Delta L = 2$ feature.\footnote{This means we concern ourselves only with Majorana neutrinos, with neutrinoless double $\beta$-decay then being an important experimental probe. For an analysis of the effective operators behind $0\nu \beta \beta$, see \cite{0nu}.}  By identifying all such independent operators, and opening each of them up in all possible ways (subject to minimality requirements), one systematically constructs radiative neutrino mass models.  The mass generation mechanism is necessarily radiative, because, unlike the ${\cal O}_1 (H^\dagger H)^n$ class, all terms in these operators contain some fields that are neither neutrinos nor neutral Higgs bosons.  The associated quanta must therefore be turned into virtual particles in loops in the neutrino self-energy diagram.  This effective operator approach is the logical extension of the Weinberg operator perspective on the see-saw mechanism.  One is simply considering models based on more complicated, and higher mass dimension, gauge-invariant $\Delta L = 2$ effective operators.

The list of SM gauge-invariant, baryon-number conserving, $\Delta L = 2$ operators formed out of quarks, leptons and the Higgs doublet has fortunately already been written down by Babu and Leung (BL) \cite{babuleung}.  We review this work in the next section, and the operator list is duplicated in the Appendix.  In Sec.~\ref{sec:loop} we investigate how to turn the effective operators into neutrino self-energy graphs by forming loops.  The next two sections, \ref{sec:4f} and \ref{sec:6f}, then provide a topological analysis of the Feynman diagrams that serve to open-up the effective operators.  Section~\ref{sec:4f} deals with operators containing four fermion fields, while the subsequent section deals with the six-fermion cases.  We restrict the exotic particles in the UV completions to scalars, vector-like Dirac fermions, and Majorana fermions.  In Sec.~\ref{sec:summary} we collect our results into a recipe of sorts, that can be used as a reference guide for those wishing to construct models from the list of effective operators.  The final section contains additional discussion and concluding remarks.

\section{The effective $\Delta L = 2$ operators}
\label{sec:opslist}

The effective operators tabulated by BL, and reproduced in the Appendix of this paper, are constructed assuming the SM gauge group, the standard quark and lepton multiplet assignments absent the right-handed neutrino, and a single Higgs doublet.  Three additional dimension-9 and twelve dimension-11 operators are obtained from combining SM dimension-4 Yukawa terms with ${\cal O}_1$ and the dimension-7 operators from this list, respectively.  Their existence was noted by BL, and explicitly written down in a later paper by de Gouv\^{e}a and Jenkins (GJ) \cite{dgj}.  They are also listed in the Appendix.  

We adopt the BL/GJ numbering scheme.  Every number corresponds to a given field content (where summing over families is understood), but many of these cases have more than one independent SU(2) index structure.  For example, ${\cal O}_3$ has two possible structures, and when we need to distinguish them we use letters from the start of the Roman alphabet, so we speak for example of ${\cal O}_{3a}$ and ${\cal O}_{3b}$, where the order is as given in the Appendix.  The operators listed explicitly in the Appendix contain only scalar and pseudoscalar Lorentz contractions.  As explained by BL, operators featuring vector, axial-vector and tensor Lorentz contractions are implicitly included in the list as well.  However, these cases are not relevant for us since we are not considering exotic spin-1 or spin-2 particles in the UV completions.  The operators in the BL list do not include SM gauge fields, which could be introduced through covariant derivatives and field-strength tensors.  Babu and Leung comment that such operators may be less interesting for neutrino mass model purposes because they may be less easily produced at tree-level from an underlying UV complete theory.  Note, though, that a recent paper discusses a 3-loop radiative neutrino mass model that reduces to an effective operator that contains $W$-boson fields in addition to right-handed charged-leptons and Higgs doublets \cite{No}.  However, the UV completion in this case is at loop-level, not tree-level, so does not provide a counter-example to the claim by BL.  \emph{A priori}, UV completions involving loops are just as valid as those at tree-level, so it may be worth revisiting effective operators containing SM gauge fields in future work.  In any case, the reader should note that our analysis does \emph{not} include models based on this class of operator.

The BL list has operators of mass dimensions 7, 9 and 11.  Dimension 13 and higher cases are (fortunately) not relevant for neutrino mass models, because they are too suppressed to be able to produce a neutrino mass as large as $0.05$ eV \cite{babuleung}.  The list is long but finite. The four dimension-7 operators are
\begin{equation}
{\cal O}_{2-4}\ \ {\rm and}\ \ {\cal O}_8,
\label{7dOp}
\end{equation}
and they all contain four fermi and one Higgs field.  There are six dimension-9 operators that contain four fermi and three Higgs fields:
\begin{equation}
{\cal O}_{5-7},\ \ {\cal O}_{61},\ \ {\cal O}_{66}\ \ {\rm and}\ \ {\cal O}_{71}.
\label{9Dfourfermi}
\end{equation}
The remaining twelve dimension-9 operators are purely six-fermi in character:
\begin{equation}
{\cal O}_{9-20}.
\label{9Dsixfermi}
\end{equation}
All fifty-two dimension-11 operators, 
\begin{equation}
{\cal O}_{21-60},\ \ {\cal O}_{62-65},\ \ {\cal O}_{67-70}\ \ {\rm and}\ \ {\cal O}_{72-75},
\label{11Dsixfermi}
\end{equation}
contain six fermi and two Higgs fields.

All of the dimension-7 and some of the dimension-9 operators have been used in the literature as bases for neutrino mass models. Only four such models have been analysed in depth so far; several others have been written down, but not fully investigated.  The historically first radiative neutrino mass model, a 1-loop scenario proposed by Zee \cite{Zee}, is based on the purely leptonic ${\cal O}_2$ operator.  The minimal Zee model is ruled out.  The operator ${\cal O}_9$ is generated in the Babu-Zee 2-loop model \cite{ZBM1,ZBM2}; this theory remains viable, though the acceptable parameter space was reduced recently from negative searches by the ATLAS and CMS collaborations \cite{SSATLAS, SSCMS}.  More recently, Babu and Julio have published detailed papers on 2-loop models associated with the dimension-7 operators ${\cal O}_3$ and ${\cal O}_8$ \cite{BJ1,BJ2}.  The following operators have received brief attention: ${\cal O}_3$ (a 1-loop variant), ${\cal O}_{4-5}$, ${\cal O}_{10-12}$, ${\cal O}_{19}$ \cite{babuleung} and ${\cal O}_{71}$ \cite{dgj}.  To the best of our knowledge, no dimension-11 operators have yet been used as the foundations of any models.

\begin{figure}
\centering
\includegraphics[height=0.20\columnwidth]{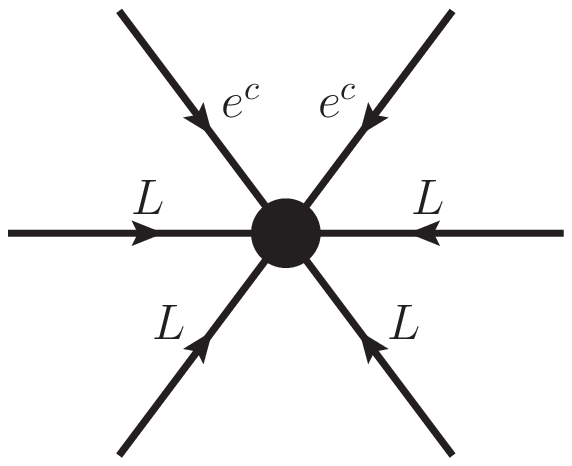}
\includegraphics[height=0.20\columnwidth]{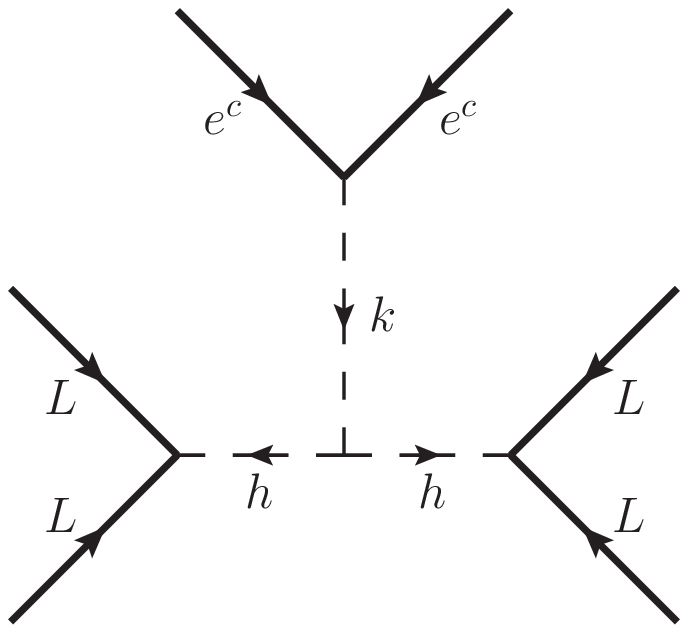}
\includegraphics[height=0.20\columnwidth]{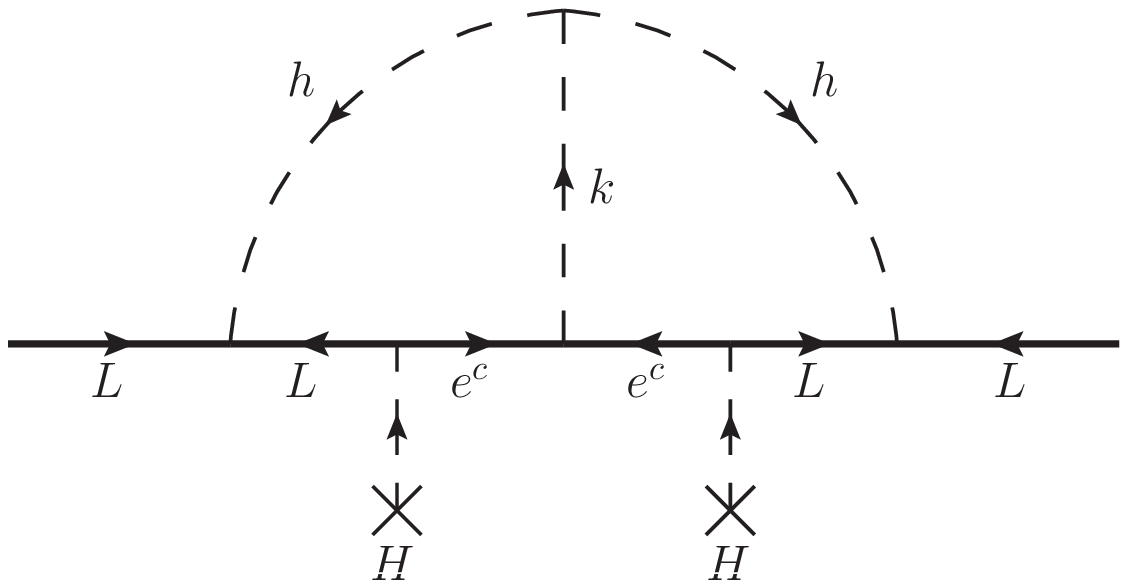}
\caption{The leftmost graph depicts the effective operator ${\cal O}_9 = LLLe^cLe^c$.  The middle graph shows how this operator can be opened up using two exotic scalars: an isosinglet, singly-charged scalar $h$ coupling to $LL$, and an isosinglet doubly-charged scalar $k$ coupling to both $e^c e^c$ and $hh$.  By forming two $L e^c$ pairs into loops via a Yukawa coupling to the Higgs VEV, the opened-up operator induces a Majorana self-energy graph for the neutrino, as shown in the rightmost plot. The result is a 2-loop contribution to the neutrino mass, with the effective operator opened-up at tree-level and the loops coming from joining external fermions via an electroweak Yukawa interaction.  This is the Babu-Zee model. Note the arrows on the fermion lines denote the flow of left-handed chirality, following the convention of BL.}
\label{fig:O9toBZ}
\end{figure}

Let us review how the Babu-Zee model can be obtained through the opening of ${\cal O}_9 = LLLe^cLe^c$.  Figure~\ref{fig:O9toBZ} summarises the procedure in diagrams -- note that in this diagram and throughout the remainder of this paper we denote the fields originating from the effective operator with bold lines.  We first note that there are two $L e^c$ pairs of external lines in ${\cal O}_9$.  Each of them can be turned into a fermion loop through a Yukawa coupling to the Higgs doublet.  When the external Higgs lines are replaced with their VEVs, the result is a 2-loop Majorana-like self-energy graph for the neutrino.  If we want a 2-loop contribution to the neutrino mass, we must therefore open-up ${\cal O}_9$ at tree-level.  The way chosen in the Babu-Zee model involves the introduction of two exotic scalars, $h$ and $k$.  They are both colourless and isosinglets, with $h$ being singly-charged and Yukawa coupling to an isosinglet $LL$ combination, and $k$ being doubly-charged.  It Yukawa couples to $e^c e^c$ and through a cubic scalar interaction also to $hh$.  The finite neutrino self-energy graph in the UV complete theory is shown in the rightmost graph of Fig.~\ref{fig:O9toBZ}.

Note that though the directions of the fermion lines in this figure appear somewhat unusual, these designations are consistent with the compact notation used to write down the operators (reviewed in the Appendix).  In this notation, following BL, the arrows represent the flow of left-handed chirality.  We adopt this convention throughout this paper as it makes it straightforward to check whether a diagram is allowed by chirality, as we discuss in Sec.~\ref{sec:4f}.

The purpose of the diagram topology analysis presented in the next two sections is to generalise this process to all operators in the list, allowing exotic vector-like Dirac fermions and Majorana fermions as well as exotic scalars in the UV completions, and making sure that \emph{all} UV completions under these assumptions are determined.  This last point ensures that no models will be missed.  The topological analysis identifies the ingredients necessary to produce a loop-level neutrino self-energy graph; it does not ensure that the resulting model works in detail, either phenomenologically or in terms of self-consistency.  The successful models will be a subset of the models implicitly defined through our diagrammatic analysis.

Let us summarise the class of models under consideration in this paper, which serves also to define what we mean by ``minimal'':
\begin{enumerate}
\item The gauge group is that of the SM, and the only imposed global symmetry is that of baryon number.
\item There is a single Higgs doublet, though inert (zero VEV) scalar doublets may be allowed in the UV completions.
\item The $\Delta L = 2$ effective operators are constructed from the single Higgs doublet and quark and lepton fields absent right-handed neutrinos.
\item The exotic particles that are to be integrated out to produce the effective operators are scalars, vector-like Dirac fermions and Majorana fermions.  We allow multiple families of such particles, if required.
\item  As explained below, we restrict our analysis to 1- and 2-loop models for radiative neutrino mass generation.
\end{enumerate}
We note explicitly that any models containing extended gauge symmetries, and thus exotic spin-1 particles, are classed as non-minimal.  Also, as discussed earlier, we do not include models based on effective operators containing SM gauge fields, which is not to say that these theories are not interesting.

\section{Radiative-neutrino-mass loop diagrams}
\label{sec:loop}

The first step in passing from effective operators to UV-complete models is to close off the additional fermi fields that will not play the role of the two external neutrinos. There are three ways this can be done:  (a) formation of a propagator, (b) mass insertion via Yukawa coupling to the Higgs field, and (c) closure via a W boson. Each will be discussed below.

To begin with if the operator contains both $\psi$ and $\overline{\psi}$, then these external fermions can be connected and replaced by a propagator.  Following the conventions in BL (reviewed in the Appendix), all of our unbarred fermi fields are left-handed, whilst the barred ones are right-handed.  Accordingly the propagator will sit between a left and right projection operator, meaning a term proportional to the internal loop momentum will appear in the numerator of the amplitude; for example if the internal loop momentum is labelled $p$, a $p^{\mu}$ appears.  At 1-loop order such terms vanish by virtue of the integrand being an odd function, but this is not true at higher orders. Consider the 2-loop case, where we label the internal momentum in the second loop by $q$. As a contribution to neutrino mass must form a scalar, the $p^{\mu}$ must be contracted to give some function of $p^2$ and $p \cdot q$ on the numerator (a coupling to the external momentum will not contribute to a mass diagram). Although the $p \cdot q$ term is odd in each momentum, it is impossible to separate them into two odd integrals due to a denominator of the form $(p+q)^2$, and so the integrand will not be odd.  This argument can be generalised to diagrams containing additional loops and so we conclude that closing off the loops in this manner will not give a vanishing contribution if we have at least two loops.

\begin{figure}
\centering
\includegraphics[height=0.20\columnwidth]{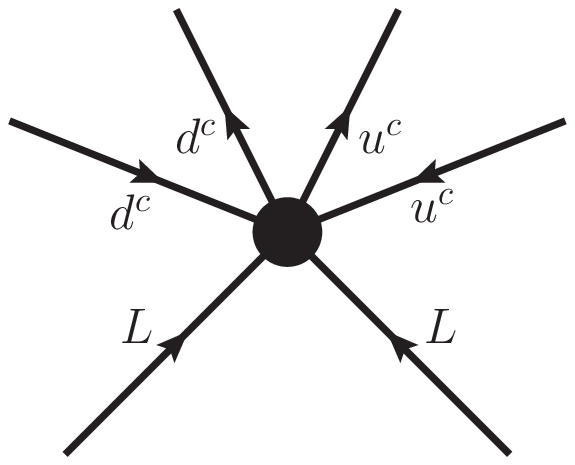}
\includegraphics[height=0.20\columnwidth]{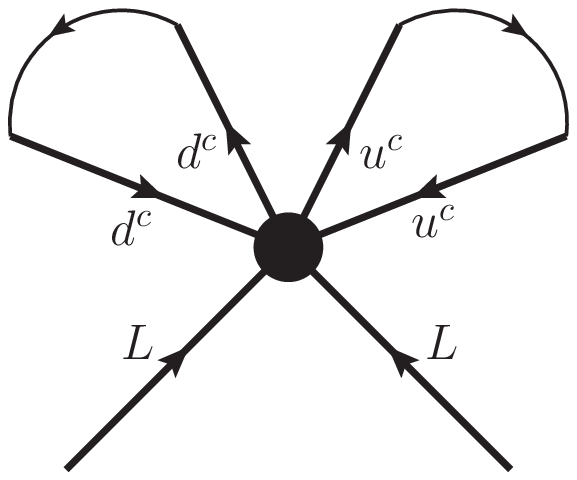}
\caption{Closing loops via the formation of a propagator for the example of ${\cal O}_{49}$.}
\label{fig:O49}
\end{figure}

An example of an operator where this procedure can be utilised is ${\cal O}_{49} = L^i L^j d^c u^c \overline{d}^c \overline{u}^c H^k H^l \epsilon_{ik} \epsilon_{jl}$; specifically we can connect $d^c$ to $\overline{d}^c$ and similarly for $u$ as depicted in Fig.~\ref{fig:O49}.  Note we have suppressed the two external Higgs lines from these diagrams -- a proper treatment of these is presented in subsequent sections.

\begin{figure}
\centering
\includegraphics[height=0.20\columnwidth]{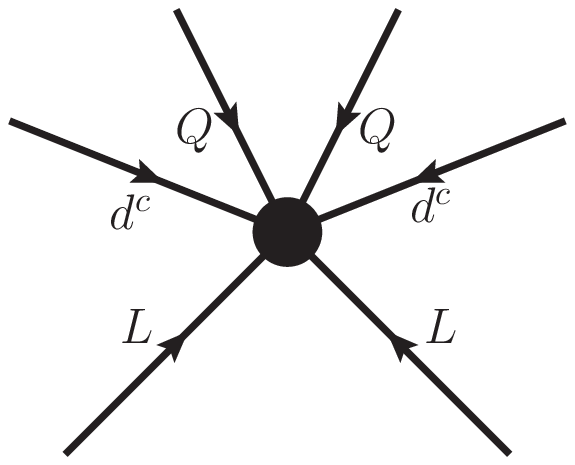}
\includegraphics[height=0.25\columnwidth]{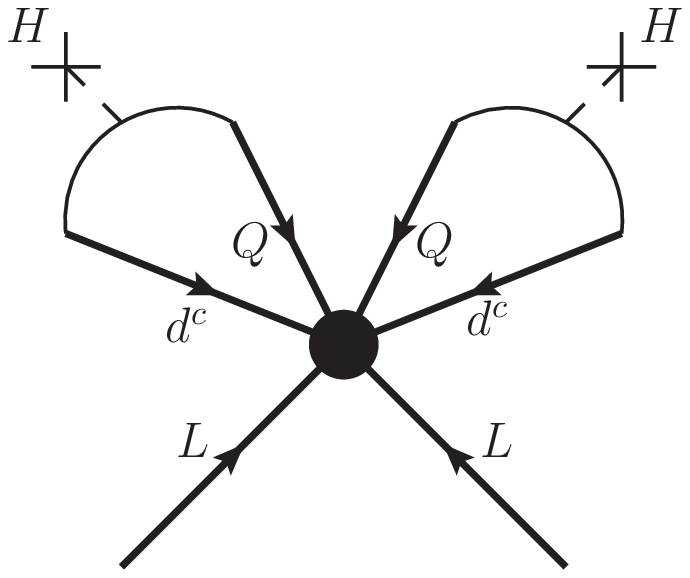}
\caption{Closing loops via a mass insertion for the example of ${\cal O}_{11b}$.}
\label{fig:O11b}
\end{figure}

Similarly if we have two fields that have an invariant coupling to the Higgs doublet, then they can be closed off via this coupling when the Higgs line is replaced by its VEV -- effectively a mass insertion, as was done in the construction of the Babu-Zee model.  A further example is furnished by ${\cal O}_{11b} = L^i L^j Q^k d^c Q^l d^c \epsilon_{ik} \epsilon_{jl}$ as seen in Fig.~\ref{fig:O11b}.

\begin{figure}
\centering
\includegraphics[height=0.20\columnwidth]{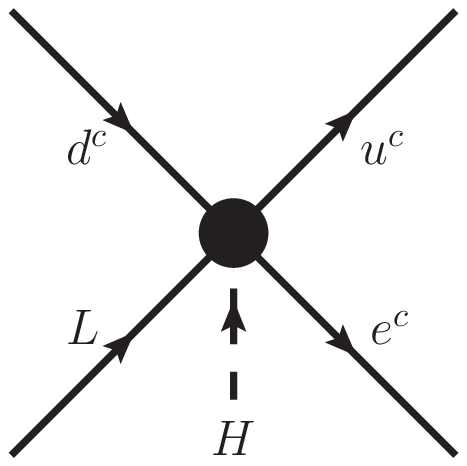}
\includegraphics[height=0.35\columnwidth]{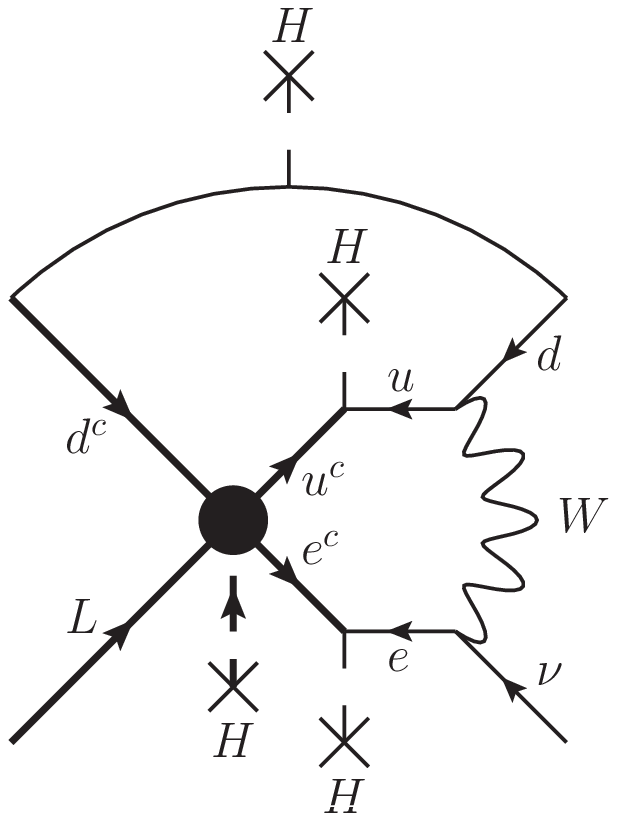}
\caption{Closing loops via insertion of a W boson for the example of ${\cal O}_8$. Note that we are working in unitary gauge, and that $e$ and $\nu$ are part of the lepton doublet $L$, whilst $u$ and $d$ are similarly from $Q$.}
\label{fig:O8}
\end{figure}

Finally there are situations where the above two options will not be available, and where closure can only be brought about by coupling to a W boson.  A simple example of this is ${\cal O}_{8} = L^i \overline{e}^c \overline{u}^c d^c H^j \epsilon_{ij}$.  There are actually two challenges in closing off this operator, as not only do the above procedures not assist in closing off the fermi fields, but also the operator does not have two external neutrinos. Both of these deficiencies can be cured by inserting a W boson and then using two mass insertions to satisfy chirality as in Fig.~\ref{fig:O8}. Note that we are here working in unitary gauge. For a general 't Hooft gauge there will be an additional diagram involving an unphysical charged Higgs.

If this final procedure is used on a six-fermion operator, then the model must contain at least three loops. As seen in the subsequent two sections, it is always possible to UV complete four- and six-fermion operators without introducing a loop in the completion.  In light of this one can catalogue the minimum number of loops required to close various operators.  One consideration that must be accounted for before doing so, however, is the SU(2) structure of the operators.  For example, many operators contain the structure $L^i L^j \epsilon_{ij} = \overline{L^c}^{~i} L^j \epsilon_{ij} = \overline{\nu^c} e - \overline{e^c} \nu$, where we have used the conventions from the Appendix.\footnote{Note that this term is nonzero only when the two $L$ fields are from different families.}  Accordingly such operators do not contain two external neutrinos without the addition of an extra loop, much like we saw for ${\cal O}_{8}$.  Similarly $Q^k H^m \epsilon_{km} = u_L H^0 - d_L H^+$, and so if this structure appears we cannot couple the $Q$ to a $d^c$ without introducing an additional loop.  Accounting for these limitations the following operators cannot be closed in less than three loops:
\begin{equation}
\begin{array}{c}
{\cal O}_{11a},\ \ {\cal O}_{12b},\ \ {\cal O}_{14a},\ \ {\cal O}_{15-20},\ \ {\cal O}_{24},\ \ {\cal O}_{26a},\ \ {\cal O}_{28},\ \ {\cal O}_{29b},\ \ {\cal O}_{30a},\ \ {\cal O}_{32}, \\
{\cal O}_{34-38},\ \ {\cal O}_{43},\ \ {\cal O}_{44c},\ \ {\cal O}_{47f},\ \ {\cal O}_{47h},\ \ {\cal O}_{50},\ \ {\cal O}_{52-60},\ \ {\cal O}_{63a},\ \ {\cal O}_{64b},\ \ {\cal O}_{65}, \\
{\cal O}_{68a},\ \ {\cal O}_{69b},\ \ {\cal O}_{70},\ \ {\cal O}_{73a},\ \ {\cal O}_{74b}\ \ {\rm and}\ \ {\cal O}_{75},
\label{threeloop}
\end{array}
\end{equation}
whilst of these the following require at least four loops:
\begin{equation}
{\cal O}_{36-38},\ \ {\cal O}_{53}\ \ {\rm and}\ \ {\cal O}_{59-60}.
\label{fourloop}
\end{equation}
As already mentioned, atmospheric and long baseline experiments are inconsistent with the neutrino acquiring its mass at 4- or higher-loop order and so we can conclude that the operators listed in Eq.~\ref{fourloop} cannot be the origin of the physical neutrino masses.  A 3-loop origin for neutrino mass does not appear to be ruled out, and indeed such models have been proposed; see for example \cite{AKS} and the aforementioned \cite{No}.  The discussion in Sec.~\ref{sec:additionalloops} should provide ample guidance for constructing 3-loop models based on the BL operator list; however, we choose to stop at two loops in the subsequent analysis, so we no longer consider the operators in Eq.~\ref{threeloop}.  Note that the list here is slightly different from that appearing in GJ, however we suspect there may have been a small error in their original list in that they assumed two loop integrals with an odd numerator vanish.

With the loops closed the remaining challenge is to UV complete the interiors. The specifics of this are covered in the following two sections.

\section{Four-fermion operators}
\label{sec:4f}

In this section we catalogue the possible UV completions for the four-fermion operators that appear in Eqs.~\ref{7dOp} and \ref{9Dfourfermi}.  To structure the discussion we consider the 1- and 2-loop cases separately and further demarcate the 1-loop case into completions involving only scalars and those with both scalars and fermions.  In the final subsection, we discuss the possibility of adding in extra loops to the minimal structures.  Recall we are working in the minimal case of the SM gauge symmetry, so we do not consider the possibility of UV completions containing new gauge bosons.  It is also worth pointing out that a recurring theme throughout this section and the next is that chirality prevents a number of operators from having certain UV completions.  This simply means it is impossible to order the fermion fields in a way that avoids a vertex containing $P_L P_R = 0$.  Using the convention outlined in Sec.~\ref{sec:opslist}, where the directions of the fermion lines denote the flow of left-handed chirality, vertices allowed by chirality must have two fermion arrows pointing in or out if they involve a scalar, or one in and one out if they involve a gauge boson.  This is the real benefit of this convention: it makes checking the chirality straightforward.

\subsection{1-loop completions}

Not all of the four-fermion operators can arise from 1-loop models.  Those that cannot are ${\cal O}_{3a}$ and ${\cal O}_{4b}$, due to their SU(2) structure, and ${\cal O}_7$ and ${\cal O}_8$, which can only be closed using two loops as in Fig.~\ref{fig:O8}.  In addition ${\cal O}_{4a}$ and ${\cal O}_6$ require both scalars and fermions in their completion to avoid chirality constraints.

\subsubsection{Scalar-only completions}

\begin{figure}
\centering
\includegraphics[height=0.20\columnwidth]{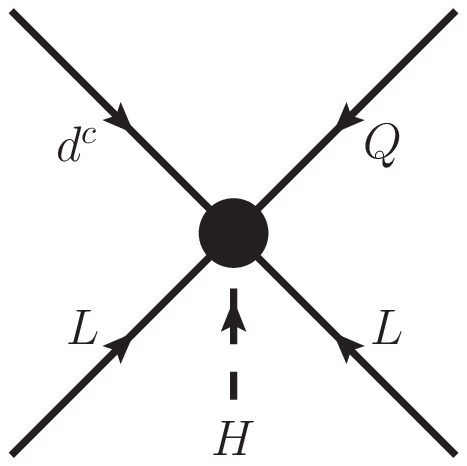}
\includegraphics[height=0.30\columnwidth]{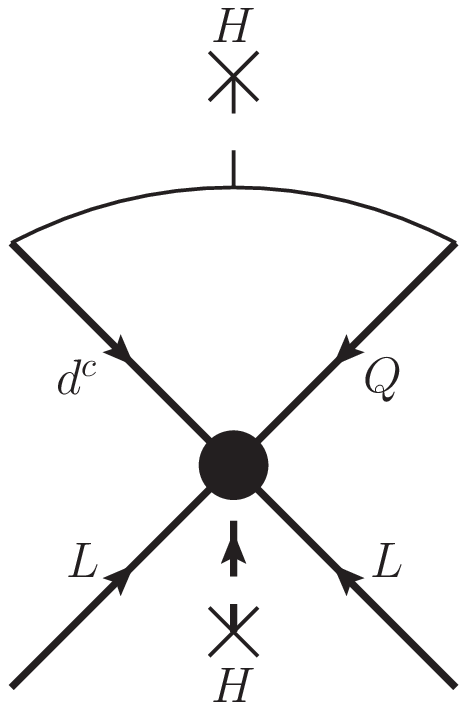}
\includegraphics[height=0.30\columnwidth]{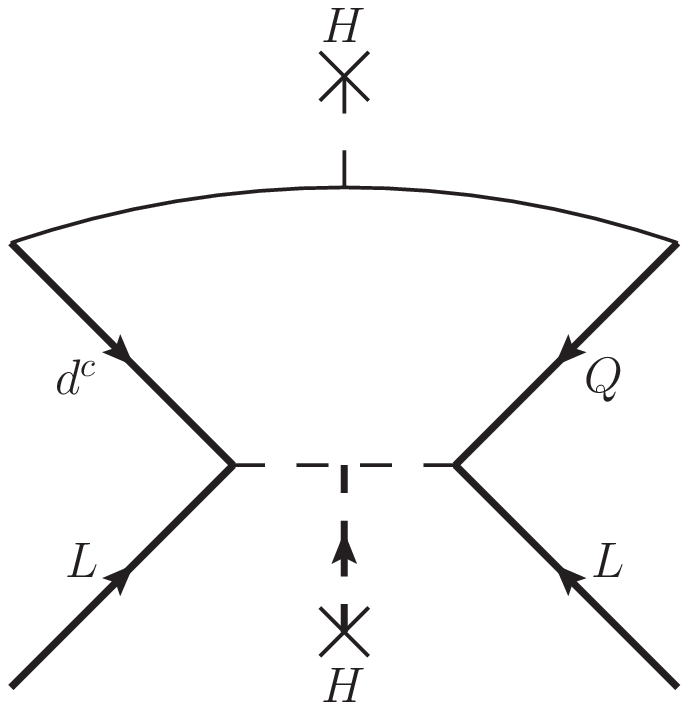}
\caption{1-loop completion of ${\cal O}_{3b}$ using scalars only.  The leftmost graph shows the effective operator vertex; in the middle we have closed the loops as in Sec.~\ref{sec:loop}; and finally on the right we have UV completed graph.  Note that we have attached the Higgs field to the new scalar to avoid introducing new fermions.}
\label{fig:scalaropen}
\end{figure}

Given that we are not considering the possibility of new gauge bosons, a renormalisable vertex with fermions must contain exactly two fermions and one scalar. Accordingly if we insist on not introducing a loop into the completion, the only way to open up operators with four fermi fields is to split them into pairs connected by a new heavy scalar.  In addition, for operators that contain a Higgs doublet, that field must be attached to this scalar line and replaced by its VEV; if it were connected to one of the SM fermion fields, this would necessarily introduce a new fermion or a SM fermion into the UV completion.  We deal with the former in the next section, but the latter is forbidden as it would mean we are no longer dealing with the same effective operator.  An example of how this procedure works is shown for ${\cal O}_{3b} = L^i L^j Q^k d^c H^l \epsilon_{ik} \epsilon_{jl}$ in Fig.~\ref{fig:scalaropen}.

The quantum numbers of the new scalars introduced in such models will be fixed -- up to a small ambiguity in the SU(3) and SU(2) values -- by the identity of the two SM fermions they connect to, from imposing gauge invariance at the vertices. Due to this, by considering which fermion couplings are allowed by chirality, it is actually possible to enumerate all the scalars such models can introduce. This is done in Table \ref{tab:newscalars}. In this table we have included all the scalars that can arise in the UV completion of four-fermion operators up to two loops, not just those that arise in the simple scalar 1-loop case. The only exception to this is that in the dimension 9 four-fermion operators, there will be new scalars that emerge from the coupling of a Higgs field to one of the scalars listed in the table. These new fields are trivially related to those listed, so we do not list them separately.

\renewcommand{\arraystretch}{1.2}
\begin{table}[htbp]
\caption{New scalars that can be introduced when UV completing four-fermion operators.  Note that redundant hermitian conjugates have been suppressed.}
\begin{center}
\begin{tabular}{l l p{10cm}}
Vertex & UV Scalar & Comment \\
\hline \hline
$\overline{L^c} L \phi$ & $\phi \sim (1,1(3),2)$ & The singlet appears in the Babu-Zee and Zee models, whilst the triplet is in the type II see-saw \\
\hline
$\overline{L^c} Q \phi$ & $\phi \sim (\bar{3},1(3),2/3)$ & - \\
\hline
$\overline{Q} u \phi$ & $\phi \sim (1(8),2,-1)$ & Singlet transforms as the SM field $\overline{H}$ \\
\hline
$\overline{Q} e \phi$ & $\phi \sim (3,2,7/3)$ & - \\
\hline
$\overline{d} L \phi$ & $\phi \sim (3,2,1/3)$ & Present in the first Babu-Julio model \\
\hline
$\overline{d} Q \phi$ & $\phi \sim (1(8),2,-1)$ & Singlet transforms as the SM field $\overline{H}$ \\
\hline
$\overline{u} L \phi$ & $\phi \sim (3,2,7/3)$ & -  \\
\hline
$\overline{e} L \phi$ & $\phi \sim (1,2,-1)$ & Transforms as the SM field $\overline{H}$ \\
\hline
$\overline{e^c} u \phi$ & $\phi \sim (\bar{3},1,2/3)$ & -
\end{tabular}
\end{center}
\label{tab:newscalars}
\end{table}

As a final comment, if the operator only contains two $L$ fields, then the $\overline{L^c} L \phi$ coupling is unfavourable.  If $\overline{L^c} L$ couples to form a singlet, then there will only be a single external neutrino and the diagram will not generate a neutrino mass.  The alternative is to couple them to form a triplet, which according to Table \ref{tab:newscalars}, implies the model will introduce the same scalar as operates to give the Type II see-saw mechanism.  This field will induce a tree-level neutrino mass, that would be expected to dominate over the 1-loop contribution, unless this lower order diagram is forbidden by a new symmetry.  In the spirit of minimality we will not be considering introducing new symmetries here, but for a comprehensive discussion on how they can be used to forbid lower order diagrams see Ref.~\cite{symforbid}.

\subsubsection{Scalar-plus-fermion completions}

\begin{figure}
\centering
\includegraphics[height=0.20\columnwidth]{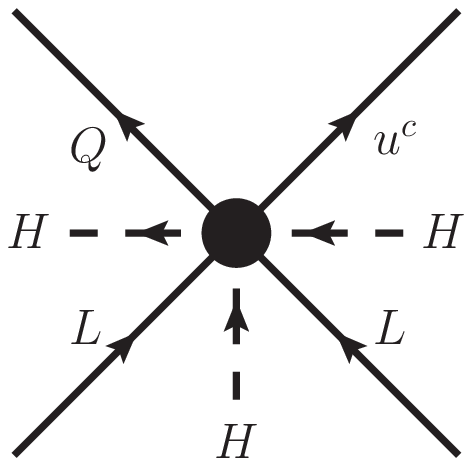}
\includegraphics[height=0.30\columnwidth]{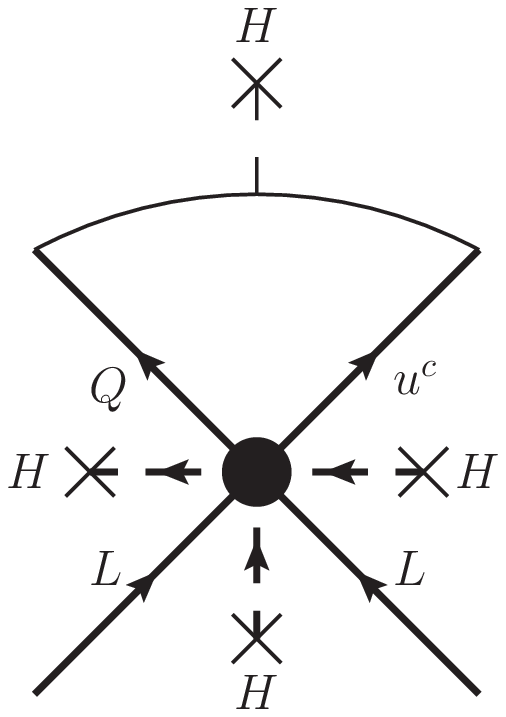}
\includegraphics[height=0.30\columnwidth]{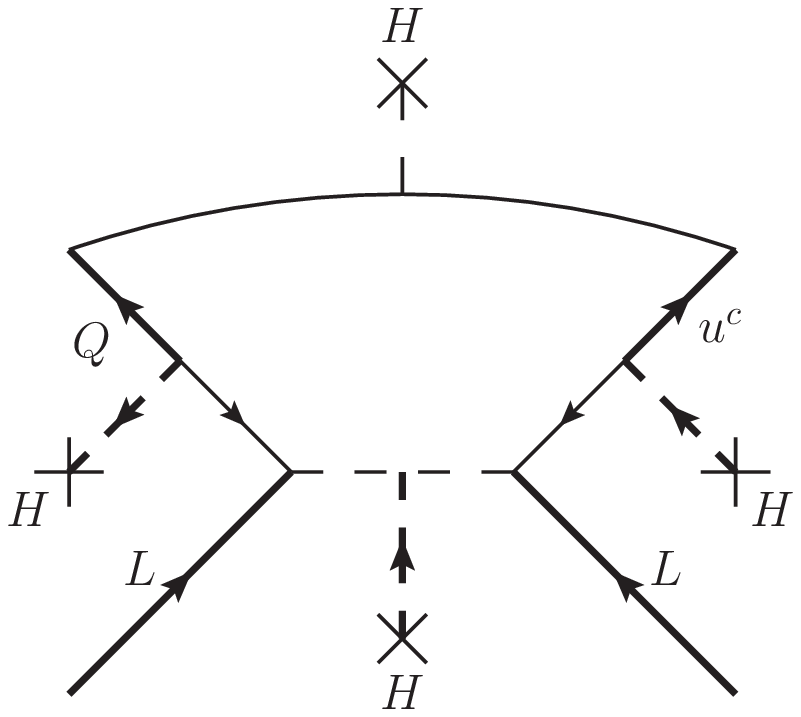}
\caption{1-loop completion of ${\cal O}_6$ using scalars and fermions.  On the left we have the effective operator; in the middle we have closed the loops; and on the right we have UV completed the central vertex.}
\label{fig:O6}
\end{figure}

Without introducing a new loop into the completion, the only way to allow new fermions into the graph is to couple the Higgs field to one of the SM fermions.  Using this procedure we can now close ${\cal O}_{4a}$ and ${\cal O}_6 = L^i L^j \overline{Q}_k \overline{u}^c H^l H^k \overline{H}_i \epsilon_{jl}$.  We show an example of completing the latter in Fig.~\ref{fig:O6}.

In Table \ref{tab:newfermions} we list all possible new fermions as we did for scalars.  Again we list all the fermions that can arise from the UV completion of four-fermion operators and we do not consider the extra possibilities from coupling a Higgs field to one of these new fermions.

\renewcommand{\arraystretch}{1.2}
\begin{table}[htbp]
\caption{New fermions that can be introduced in the UV completion.  Again we have suppressed the hermitian conjugate cases.}
\begin{center}
\begin{tabular}{l l p{10cm}}
Vertex & UV Fermion & Comment \\
\hline \hline
$\overline{f} L H$ & $f_R \sim (1,1(3),0)$ & The singlet and triplet appear in the Type I and III see-saws respectively \\
\hline
$\overline{f} L \overline{H}$ & $f_R \sim (1,1(3),-2)$ & The singlet transforms as the SM field $e_R$ \\
\hline
$\overline{f} Q H$ & $f_R \sim (3,1(3),4/3)$ & The singlet transforms as the SM field $u_R$ \\
\hline
$\overline{f} Q \overline{H}$ & $f_R \sim (3,1(3),-2/3)$ & The singlet transforms as the SM field $d_R$ \\
\hline
$\overline{e} f H$ & $f_L \sim (1,2,-3)$ & - \\
\hline
$\overline{e} f \overline{H}$ & $f_L \sim (1,2,-1)$ & Transforms as the SM field $L_L$ \\
\hline
$\overline{u} f H$ & $f_L \sim (3,2,1/3)$ & Transforms as the SM field $Q_L$ \\
\hline
$\overline{u} f \overline{H}$ & $f_L \sim (3,2,7/3)$ & - \\
\hline
$\overline{d} f H$ & $f_L \sim (3,2,-5/3)$ & - \\
\hline
$\overline{d} f \overline{H}$ & $f_L \sim (3,2,1/3)$ & Transforms as the SM field $Q_L$
\end{tabular}
\end{center}
\label{tab:newfermions}
\end{table}

We have already noted that we will only be considering adding vector-like Dirac fermions or Majorana fermions to the UV completion. This is for the pragmatic reason that their masses can then be decoupled from the electroweak scale and thereby avoid any tension from their experimental non-observation. Nonetheless in several cases chirality mandates that vector-like fermions be used. This is the case if we want a 1-loop model with a new fermion from ${\cal O}_{3b}$ and ${\cal O}_{4a}$. These operators only have a single Higgs field that can be used to create a new fermion, however once this is inserted chirality forces the diagram to vanish. A solution would appear to be making the new fermion a Majorana particle, as then a Majorana mass term can be introduced to give a further chirality flip. Looking at Table \ref{tab:newfermions}, the only possibility is $f_R \sim (1,1(3),0)$ as a Majorana fermion must have vanishing hypercharge. Yet these are exactly the fermions that appear in the Type I and III see-saw mechanisms, so the putative 1-loop models would actually induce a dominant tree-level contribution, thus defeating the purpose of the model.  As such, the only possibility is to make the new fermion vector-like, as then the required chirality flip can be furnished by the mass term $m_f \overline{f}_L f_R$ or its conjugate.

\subsection{2-loop completions}

\begin{figure}
\centering
\includegraphics[height=0.35\columnwidth]{O8after.eps}
\includegraphics[height=0.35\columnwidth]{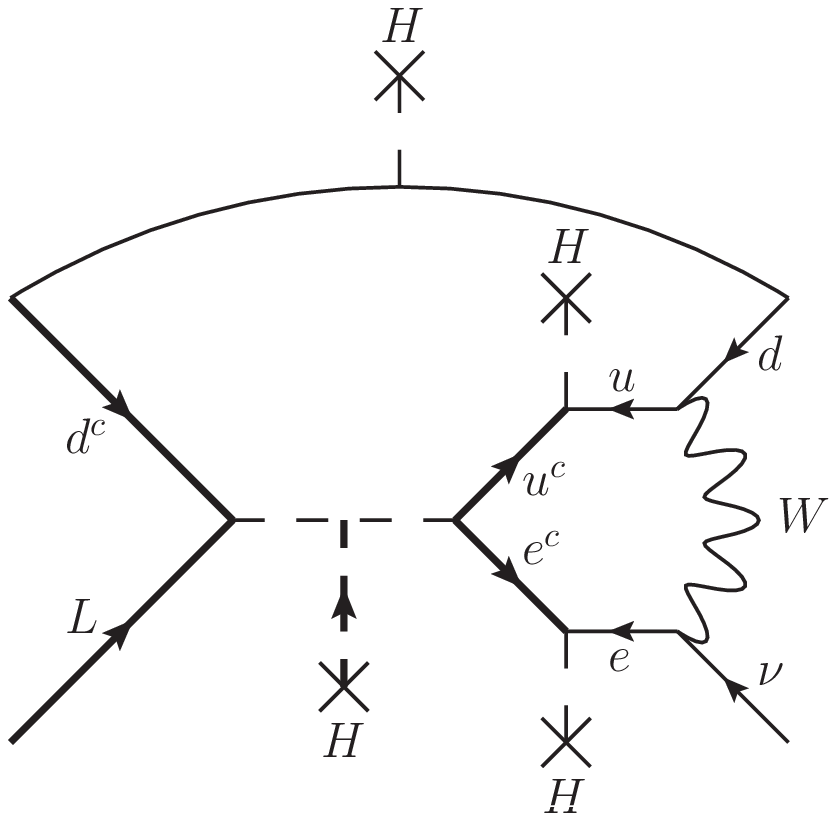}
\caption{Closure of ${\cal O}_8$ (left) and its UV completion (right).}
\label{fig:O8UV}
\end{figure}

As mentioned, ${\cal O}_{3a}$, ${\cal O}_{4b}$, ${\cal O}_{7}$ and ${\cal O}_{8}$ can only be closed in two loops.  We have already outlined how to close the loops for ${\cal O}_{8}$ in Fig.~\ref{fig:O8}.  From here the UV completion is straightforward and there is a single possibility as shown in Fig.~\ref{fig:O8UV} -- the other possible option of placing a scalar between $L \overline{e}^c$ and $\overline{u}^c d^c$ vanishes as both vertices are forbidden by chirality.  The Higgs field has been arbitrarily attached to the new scalar line; its placement on a fermion line dictates the new fermion is vector-like according to the above discussion. The completions of ${\cal O}_{3a}$, ${\cal O}_{4b}$ and ${\cal O}_{7}$ are analogous.

\subsection{Additional Loops}
\label{sec:additionalloops}

\begin{figure}
\centering
\includegraphics[height=0.30\columnwidth]{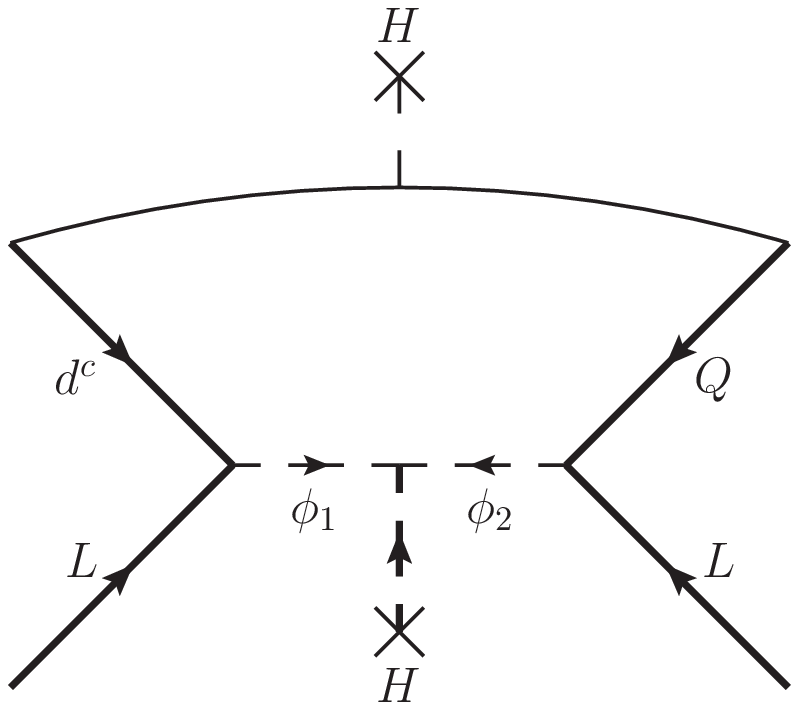}
\includegraphics[height=0.30\columnwidth]{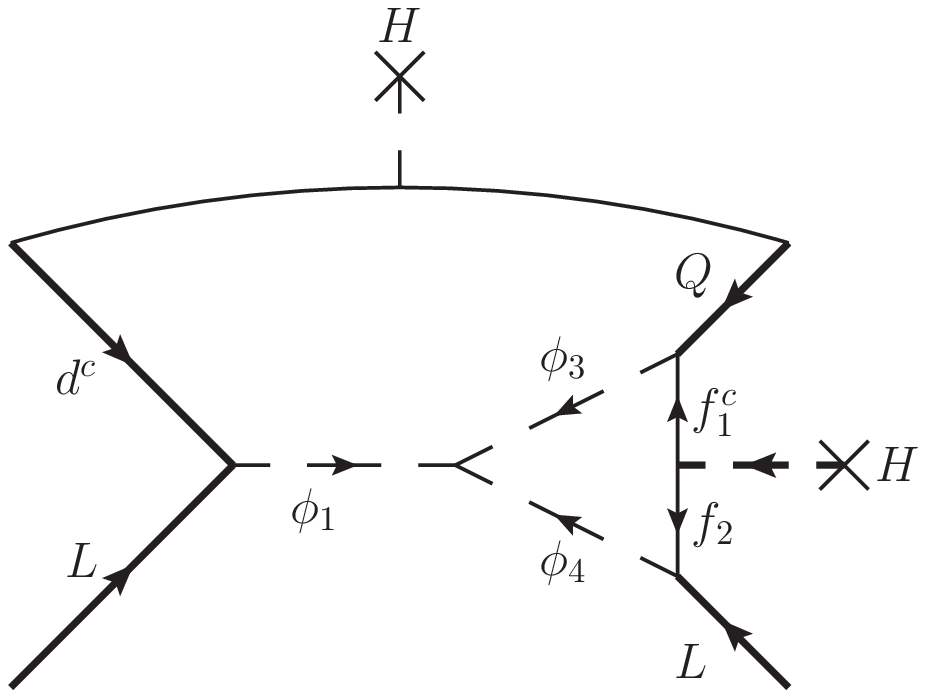}
\caption{1- and 2-loop graphs from ${\cal O}_{3b}$.}
\label{fig:O3twoloop}
\end{figure}

So far we have focussed on models with the fewest possible loops that can be derived from four-fermion operators.  In general it is possible to add extra loops to these structures. A simple possibility is to add loops into the UV completion.  We show an example of how this can be done for ${\cal O}_{3b}$ in Fig.~\ref{fig:O3twoloop}, where all new fields have been labelled distinctly.  Note that this is just one of a number of different ways a loop can be added into the UV completion. With the Higgs fields positioned as shown in the diagram, $\phi_2$ appears in the 1-loop but not the 2-loop case. For this reason the 2-loop diagram will not induce the 1-loop diagram, making it a potentially interesting model.  However, one can show that the quantum numbers for the new fields in the 2-loop diagram are not fixed uniquely: there are an infinite number of forms the new particles can take.  This is a generic feature of adding additional loops to the UV completion.  So, whilst it is always possible to add additional loops in this way to the four- and six-fermion models we describe, we will not be considering these somewhat ill-defined models in any more detail in this paper.

\begin{figure}
\centering
\includegraphics[height=0.30\columnwidth]{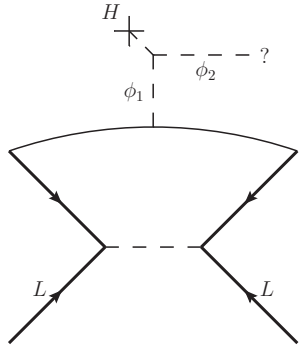}
\caption{Adding an external loop to create a 2-loop model from four-fermion operators. There is no unique position for $\phi_2$ to connect.}
\label{fig:ExternalLoop}
\end{figure}

The alternative is to add external loops to the existing structure. If this is done using SM fields, then the lower order structure will always be present.  Thus these diagrams are irrelevant from the perspective of neutrino mass.  Nevertheless in the case of a four-fermion operator where the loops are closed through a mass insertion via Yukawa coupling to the Higgs field, there is a non-degenerate way to add an external loop. Firstly if we have a $\psi$ and $\chi$ that have an invariant coupling $\overline{\psi} \chi H$ or $\overline{\psi} \chi \overline{H}$, then we can introduce an inert (i.e.\ zero VEV) Higgs-like scalar $\phi_1$ to replace the $H$ in these couplings.\footnote{To prevent $\phi_1$ obtaining an induced VEV, terms such as $\bar{H} \phi_1$ must be forbidden. If $\phi_1$ couples to $\overline{Q} u$, $\overline{Q} d$ or their conjugates, this can be ensured by choosing $\phi_1$ to transform as an octet under SU(3). If $\phi_1$ couples to $\overline{L} e$ or its conjugate, then an induced VEV can only be forbidden by an imposed discrete symmetry.} Next, note that ensuring $\phi_1$ does not acquire a VEV is not sufficient to prevent a 1-loop coupling. This is because if we simply close off the $\phi_1$ loop by connecting it somewhere else on the diagram, then at both points where $\phi_1$ connects there will also be an allowed coupling to the Higgs field, which can be replaced by a Higgs VEV. Thus this closure alone will always induce a dominant 1-loop contribution. Nevertheless if we introduce an additional new scalar through the cubic scalar interaction $\phi_1 \phi_2 H$ or $\phi_1 \phi_2 \overline{H}$, then connecting $\phi_2$ back into the diagram will create an irreducible 2-loop graph. The exact position where $\phi_2$ attaches is not fixed; it can either be to the existing new scalar line or to a SM fermion. The latter option will introduce a new UV fermion and requires careful consideration of the chirality. Depending on where $\phi_2$ is attached, there can arise new fields to those listed in the tables above. Nonetheless these will be obviously related to those we have introduced, so we have not reproduced them here. The general setup is shown in Fig.~\ref{fig:ExternalLoop}, and a specific example that arises in the discussion of Sec.~\ref{sec:conc} is shown on the right of Fig.~\ref{fig:O68b}.

\section{Six-fermion operators}
\label{sec:6f}

The vast majority of neutrino mass effective operators contain six fermi fields, and as seen in Sec.~\ref{sec:loop}, UV-complete models associated with these feature a minimum of two loops for the neutrino self-energy graph.  These operators are listed in Eqs.~\ref{9Dsixfermi} and \ref{11Dsixfermi}.  Again it is possible to UV complete these operators using only scalars, or with both fermions and scalars, and we discuss these cases separately below.

\subsection{Scalar-only completions}

\begin{figure}
\centering
\includegraphics[height=0.20\columnwidth]{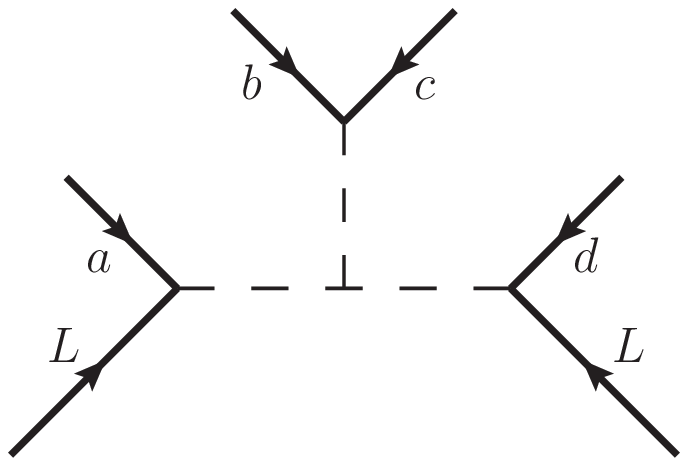}
\includegraphics[height=0.20\columnwidth]{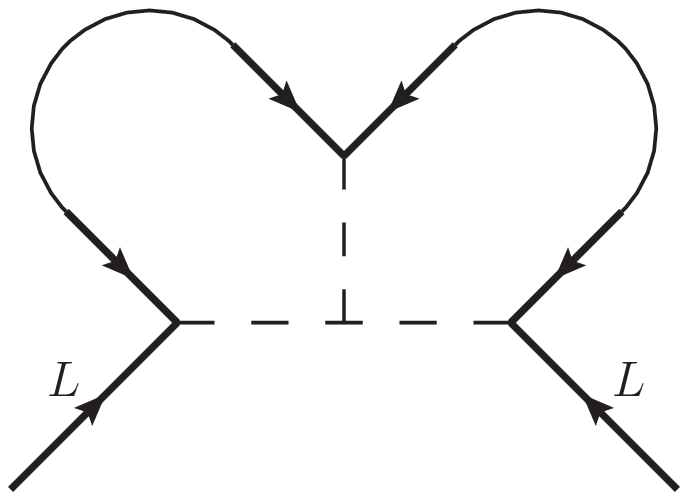}
\caption{Scalar UV completion of six-fermion operators (left) and their unique loop closure (right).  The closures on the right are a shorthand for the three options from Sec.~\ref{sec:loop}.  Note that closure via a W boson insertion would introduce an additional external loop not shown.}
\label{fig:6fscalar}
\end{figure}

If we insist on introducing only new scalars, then the only possible UV completion is to split up the six fermions into pairs connected by these scalars.  Exactly this setup is shown on the left of Fig.~\ref{fig:6fscalar}, where we have labelled each of the fermions to aid the discussion.  The position of the two $L$ fields is mandated by our earlier comment: if they appear at the same vertex we will only have one external neutrino or an induced tree-level contribution from the Type II see-saw mechanism.  Although we have already discussed how to close off extra fermion lines in Sec.~\ref{sec:loop}, the issue here is whether we do so by connecting 1) $a$ to $b$ and $c$ to $d$, or 2) $a$ to $d$ and $b$ to $c$ (the remaining permutation is topologically equivalent to the first).  There is nothing wrong with the first of these and we have displayed this on the right of Fig.~\ref{fig:6fscalar} using an obvious shorthand for the closure.  The second option, however, is not allowed.  In such a diagram there will be a fermion loop connected to the rest of the diagram only by a single scalar line.  The diagram is not 1-particle irreducible, in other words, and furthermore the 1-loop subgraphs are divergent.  Such diagrams are obviously irrelevant in the study of radiative neutrino mass models.

The basic scalar completion of six-fermion operators allows additional new scalars that were not available in the four-fermion case. These are listed in Table \ref{tab:6fermionnewscalars}.

\renewcommand{\arraystretch}{1.2}
\begin{table}[htbp]
\caption{Additional new scalars allowed for six-fermion models.}
\begin{center}
\begin{tabular}{l l p{8cm}}
Vertex & UV Scalar & Comment \\
\hline \hline
$\overline{Q^c} Q \phi$ & $\phi \sim (3(\bar{6}),1(3),-2/3)$ & - \\
\hline
$\overline{e^c} e \phi$ & $\phi \sim (1,1,4)$ & Present in the Babu-Zee model \\
\hline
$\overline{d^c} d \phi$ & $\phi \sim (3(\bar{6}),1,4/3)$ & - \\
\hline
$\overline{u^c} u \phi$ & $\phi \sim (3(\bar{6}),1,-8/3)$ & - \\
\hline
$\overline{e^c} d \phi$ & $\phi \sim (\bar{3},1,8/3)$ & - \\
\hline
$\overline{u^c} d \phi$ & $\phi \sim (3(\bar{6}),1,-2/3)$ & -
\end{tabular}
\end{center}
\label{tab:6fermionnewscalars}
\end{table}

\begin{figure}
\centering
\includegraphics[height=0.20\columnwidth]{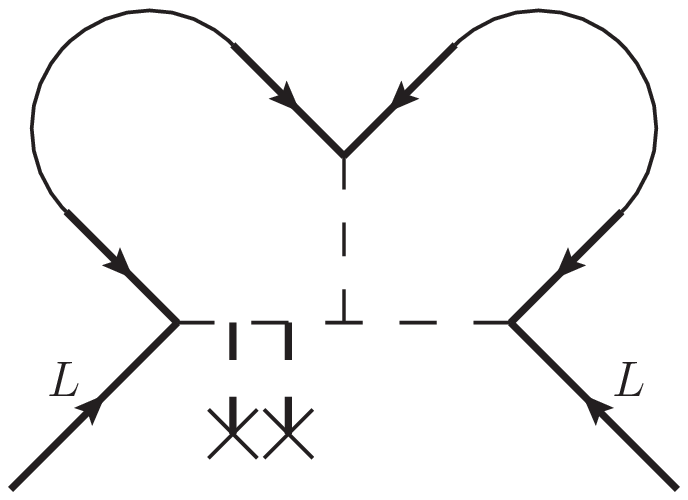}
\includegraphics[height=0.20\columnwidth]{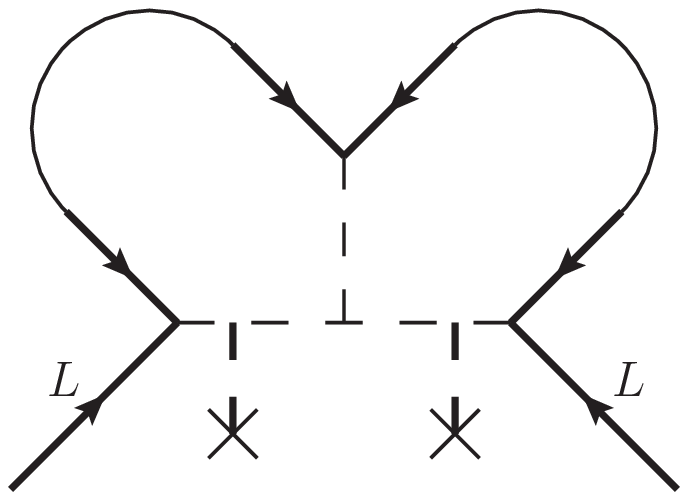}
\includegraphics[height=0.20\columnwidth]{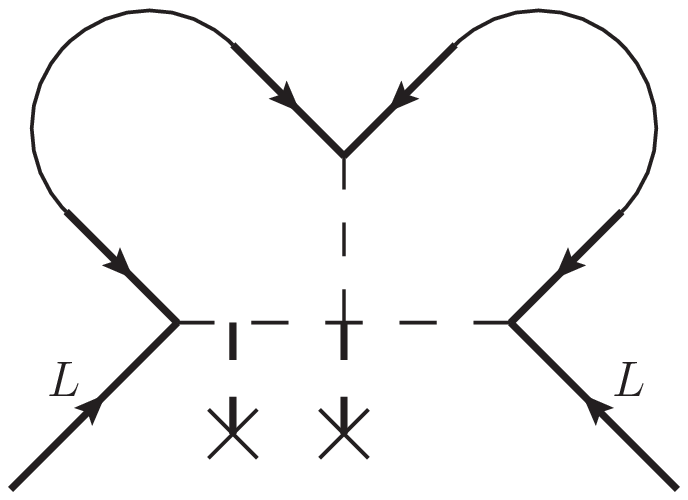} \\
\includegraphics[height=0.20\columnwidth]{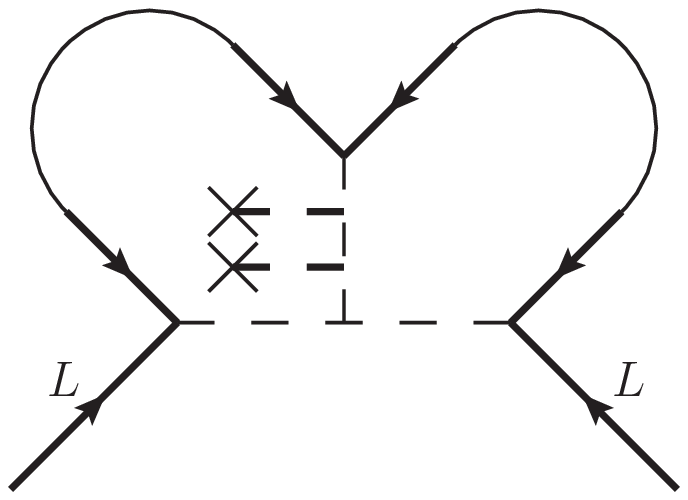}
\includegraphics[height=0.20\columnwidth]{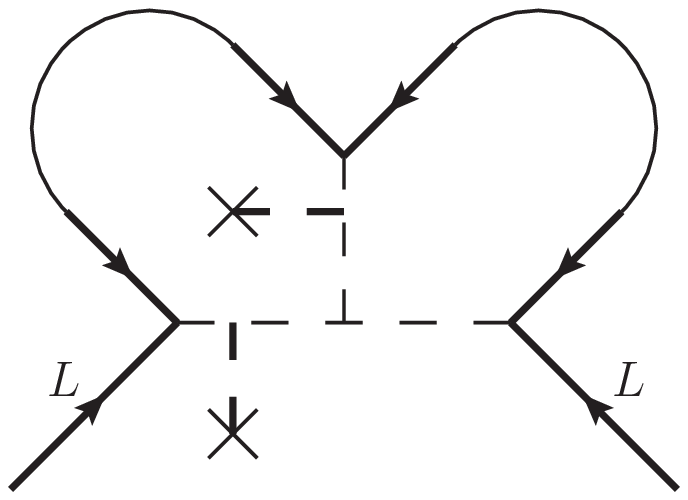}
\includegraphics[height=0.20\columnwidth]{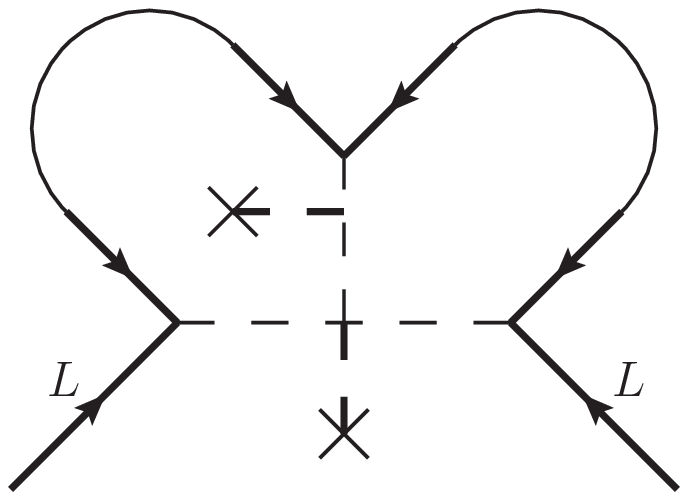}
\caption{Allowable ways to attach Higgs fields onto the six-fermion scalar only UV completion. Note there is a variation of the two leftmost graphs not shown, which is where the two Higgs or Higgs anti-Higgs fields are placed at the same position.}
\label{fig:6fscalarHiggs}
\end{figure}

Lastly we consider the possible placement of the two Higgs fields that appear in the 11D effective operators.  Insisting on introducing only scalars into the UV completion, it is apparent that the two Higgs or the Higgs anti-Higgs fields must be attached to the new scalars and then replaced by their VEVs. Despite this there are still eight topologically distinct placements, six of which we show in Fig.~\ref{fig:6fscalarHiggs}.  The two cases suppressed are when the Higgs fields are at the same location on the scalar line, which is similar to the two leftmost diagrams in the figure.  Interestingly this case is always present when we attach the Higgs fields on the same scalar line, even if not at the same point.  To see this say we have the couplings $\phi_1 \phi_2 H$ and $\phi_3 \overline{\phi_2} H$, where $\phi$ denotes a new heavy scalar.  Then by gauge invariance, as $\phi_1 H$ must transform as $\overline{\phi_2}$, this setup will always imply an invariant coupling $\phi_1 \phi_3 HH$ - the case where the two Higgs fields are at the same point.  An identical argument would hold if we have a pair of Higgs and anti-Higgs fields, but note we cannot run the argument in reverse as $\phi_2$ might not exist elsewhere in the model.  In terms of the impact these setups will have on the amplitude calculated from these diagrams, if a Higgs field is placed exactly at the vertex of the three scalars, then this will simply change the dimensionless quartic coupling constant $\lambda$ to the mass-dimension-one cubic coupling $\lambda v$.  Alternatively if the Higgs fields are attached directly on the scalar propagators, then expanding around their VEVs leads to a mixing between the scalars on the line.  In this case the scalars can be replaced by their mass eigenstates with a mixing matrix appearing in their interactions.  The technical details of this replacement have been calculated in \cite{BJ1}.

\subsection{Scalar-plus-fermion completions}

There are two ways to add fermions into the UV completion of six-fermion operators: take the scalar UV completion and attach a Higgs field to a SM fermion, or use a UV completion that introduces fermions in a topologically-different way.  We will discuss these cases separately.  Before doing so, however, there are two recurring points in the subsequent analysis that are worth emphasising at the outset.  First, whilst there is a large class of possible models once fermions are included in the UV completion, these are not all allowable for 9D operators, whilst they are for their 11D counterparts.  Second, the class of models is large enough that it would be impractical to list all the new fermions introduced; it is possible to get a fermion with almost any combination of the following quantum numbers: SU(3) $\in \{1,3,\bar{3},6,\bar{6},8\}$, SU(2) $\in \{1,2,3,4\}$ and $Y \in \{-18/3,-17/3,...,18/3\}$.

\subsubsection{Adding fermions to the scalar UV completion}

The idea here is to take the scalar UV completion structure discussed above and introduce fermions by attaching Higgs fields to the SM fermions, thereby introducing new heavy fermions.  The process is analogous to how we introduced fermions in the four fermi operator case.  Clearly this process is dependent on the effective operator containing Higgs fields and thus is only relevant for 11D operators.

Next observe that all effective operators contain an even number of left- and right-handed operators and that the operators are structured such that closing the loops as described in Sec.~\ref{sec:loop} can only bring about an even number of chirality flips.  In addition the scalar UV completion requires the coupling of three pairs of like-chirality fermions to ensure the diagram does not vanish.  As such introducing new fermions and thus chirality flips can only be done if the number of flips introduced is even.  This can be done by attaching two Higgs fields to the SM fermion lines, or alternatively using one Higgs field in conjunction with a new vector-like fermion, as we get an extra chirality flip from the mass term. In the latter case the remaining Higgs field can be attached to one of the scalar lines. Bearing such considerations in mind, it is then simple enough to write down all allowed positions of the Higgs fields in the spirit of the examples shown in Fig.~\ref{fig:6fscalarHiggs} for the scalar only case. Although there can be a large number of them for a given operator, writing these down systematically is trivial and so we have not presented them here.

\begin{figure}
\centering
\includegraphics[height=0.30\columnwidth]{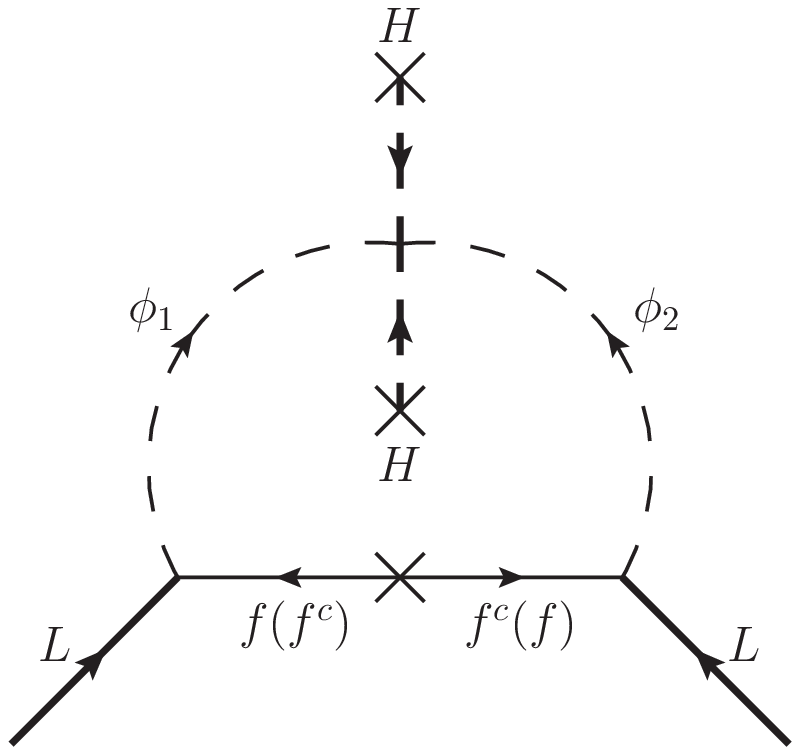}
\includegraphics[height=0.30\columnwidth]{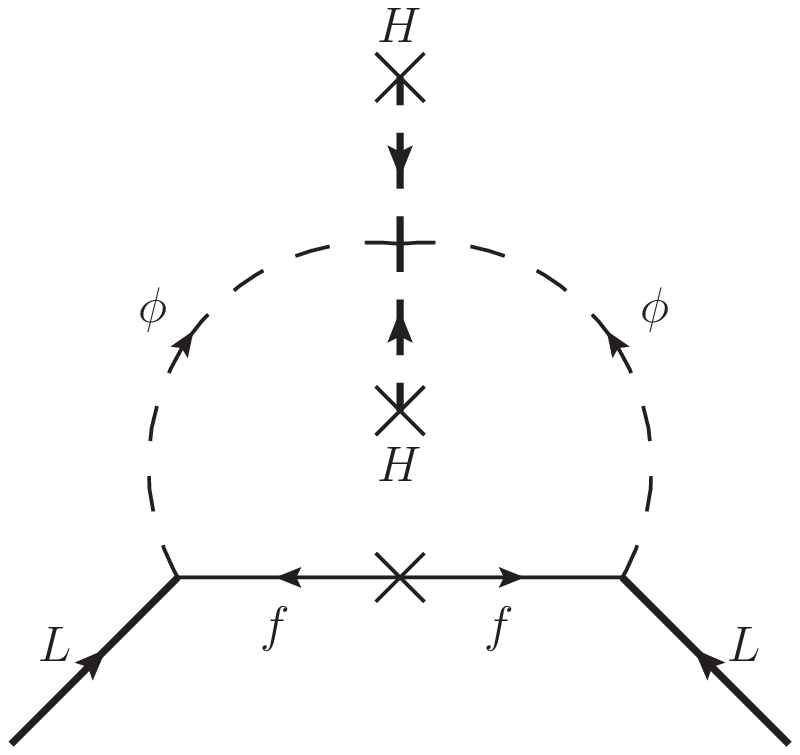}
\caption{Induced 1-loop graph in the vector-like (left) and Majorana (right) fermion case. Both of these can be integrated back to ${\cal O}_1$.}
\label{fig:inducedoneloop}
\end{figure}

As already mentioned, our analysis is only for the generation of neutrino mass diagrams, and whether these diagrams are associated with a viable model is a separate question we are not considering in detail. Nevertheless we will here give a flavour of what can go wrong, as we will need to make use of this result in the following section. Consider introducing a new vector-like fermion that couples to both $L$ and $L^c$ and a new heavy scalar at each vertex, say $\phi_1$ and $\phi_2$ respectively. Then these two vertices ensure an additional coupling will be gauge invariant -- $\phi_1 \phi_2 H H$ -- and this is enough to induce the 1-loop diagram seen on the left of Fig.~\ref{fig:inducedoneloop}. This diagram originates from ${\cal O}_1$ and will dominate over any 2-loop graph, meaning the original combination should be avoided in order to generate valid 2-loop models. As a special case, if the fermion is a Majorana particle, then simply the coupling to $L$ and a new scalar $\phi$ is sufficient to generate the diagram on the right of Fig.~\ref{fig:inducedoneloop}, which can again be integrated back to ${\cal O}_1$. In general 1-loop contributions can arise in a number of other ways, and this is a necessary consistency check for models.

\subsubsection{Central fermion in the UV completion}

\begin{figure}
\centering
\includegraphics[height=0.20\columnwidth]{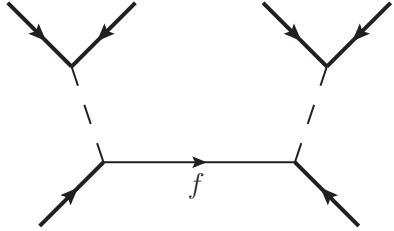}
\caption{Central fermion UV completion of six-fermion operators.}
\label{fig:6ffermion}
\end{figure}

\begin{figure}[h]
\centering
\subfigure[\ Diagram A (left), B (centre) and C (right)]{
  \includegraphics[height=0.20\columnwidth]{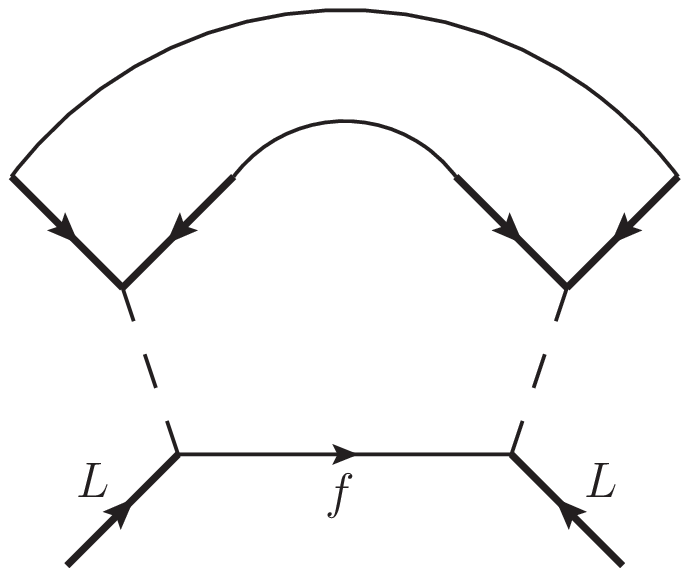}
  \includegraphics[height=0.20\columnwidth]{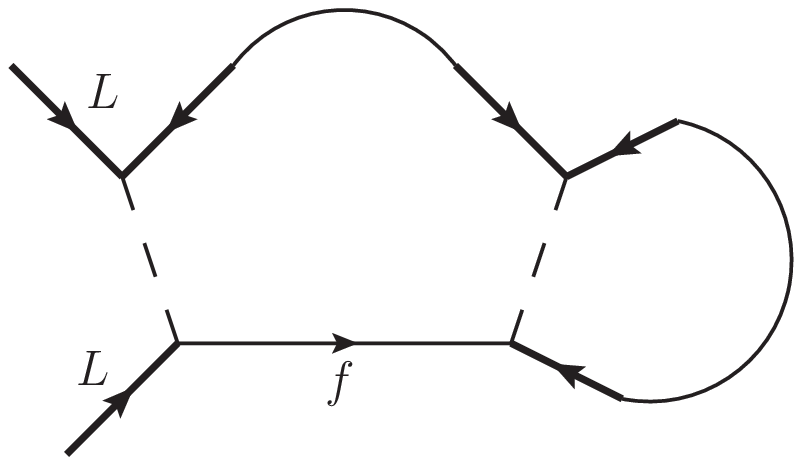}
  \includegraphics[height=0.20\columnwidth]{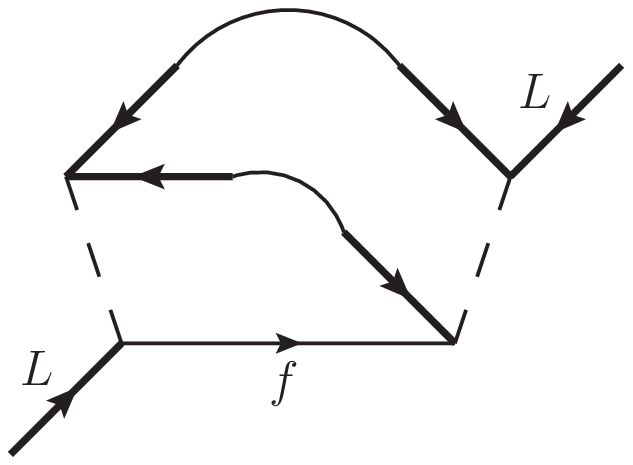}
}
\subfigure[\ Diagram D1 (left) and D2 (right)]{
  \includegraphics[height=0.20\columnwidth]{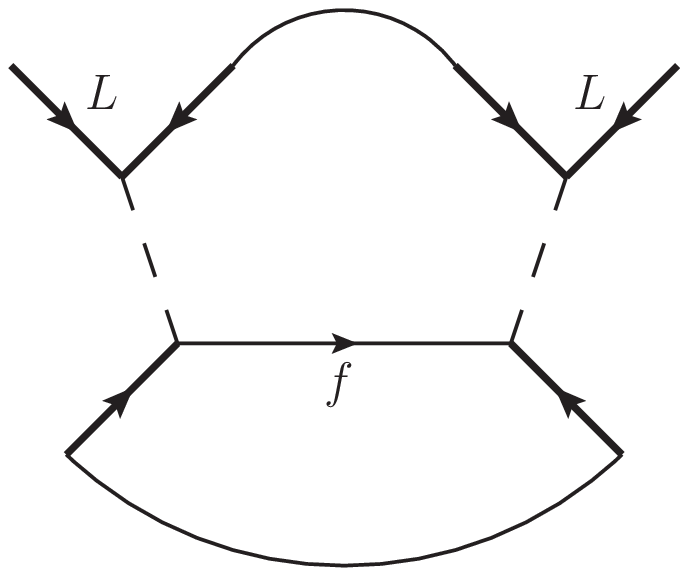}
  \includegraphics[height=0.20\columnwidth]{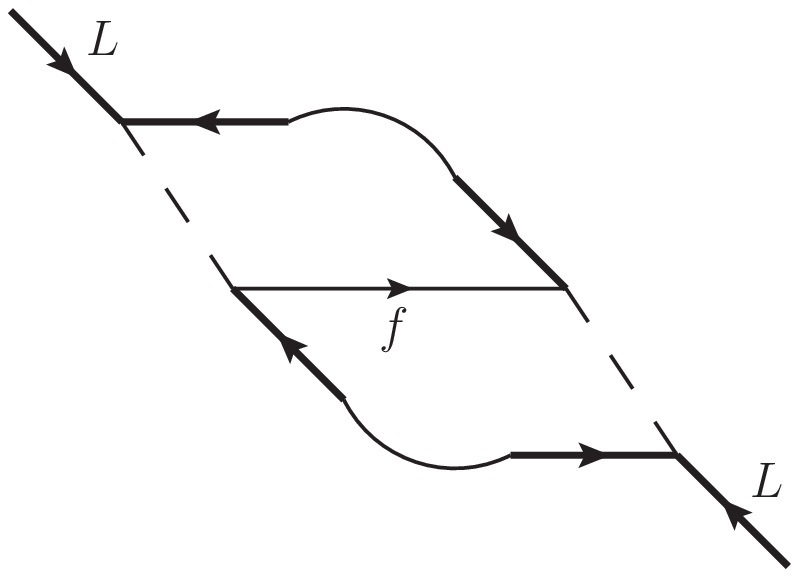}
}
\caption{\label{fig:6ffermionclosure} Central Fermion UV completions. The same loop closure shorthand as Fig.~\ref{fig:6fscalar} is employed.}
\end{figure}

Without considering loops in the UV completion, including fermions allows a single additional UV completion to that seen in Fig.~\ref{fig:6fscalar}.  This structure, which involves a new heavy central fermion, is displayed in Fig.~\ref{fig:6ffermion}.  At this stage we have not shown the placement of $LL$ explicitly, as there are four allowable placements that avoid two $L$ fields coupling at the same vertex.  Three of these four have a unique loop closure, whilst the fourth has two possibilities.  All of these are depicted in Fig.~\ref{fig:6ffermionclosure}.

Although this is a large number of possible new diagrams, not all can be constructed from 9D operators. In order to see this observe that as written all the diagrams in Fig.~\ref{fig:6ffermionclosure} are forbidden by chirality.  Accordingly to avoid the fatal $P_L P_R$ coupling we must introduce an additional chirality flip.  As 9D operators contain no Higgs fields, the only possibility is for the new heavy central fermion in the UV completion to be a vector-like fermion, as its mass insertion can provide the required chirality flip.  Nevertheless in the case of diagram A, it will be coupled to an $L$ and $L^c$ field, as well as two new scalars. This is sufficient to generate the 1-loop on the left of Fig.~\ref{fig:inducedoneloop}, which will clearly make the 2-loop contribution redundant. Furthermore if the central fermion is a Majorana particle, then the graph on the right of Fig.~\ref{fig:inducedoneloop} will be induced for diagram A, B or C. Thus we conclude the 9D operators can only make use of the central fermion UV completion in the case of diagram B, C or D if the fermion is vector-like, and only D if it is a Majorana particle.

As attaching a Higgs field to a fermion line will introduce an additional new UV fermion and chirality flip, this restriction does not apply to 11D models. Indeed given the numerous ways Higgs fields can be validly attached into the different diagrams in Fig.~\ref{fig:6ffermionclosure}, the space of allowable diagrams for 11D operators appears to be far larger than for their lower dimensional counterparts.

\section{A recipe for model building}
\label{sec:summary}

In the spirit of the presentation of the operator analysis in GJ, we have collected our final results in Table \ref{tab:summary}. We only list those operators that can be closed in two loops or less. Between this table and the various figures referred to, one should easily be able to construct all 1- and 2-loop models from a given operator.  In the table, as well as listing the appropriate loop closure technique and available topologies, we have also reproduced the inferred upper bound on the scale of new physics $\Lambda_{\nu}$. These values were derived in GJ by equating an approximate form of the neutrino mass expression to the atmospheric limit of 0.05 eV, and then extracting $\Lambda_{\nu}$ under the assumption that all of the new dimensionless coupling constants were of order one.  Because of this last assumption, the derived $\Lambda_{\nu}$ is an approximate \emph{upper limit} on the scale of new physics allowable in these operators. The scale will be lower, and can be brought into the LHC regime, by having coupling constants that are smaller than one.\footnote{We note remarks already made in Sec.~\ref{sec:intro}, that many such coupling constants will have to be less than one.}  We have also updated $\Lambda_{\nu}$ values where the number of loops the operator can be closed in has been altered, as discussed in Sec.~\ref{sec:loop}. Finally we have not included details of where Higgs fields, if present, can be located. There are several comments on this in the above sections, but in general the placement of a Higgs field is only weakly constrained -- there will be a number of allowable placements. As such models involving Higgs fields will in general give rise to significantly more diagrams than those without them.

\renewcommand{\arraystretch}{1.0}
\LTcapwidth= \textwidth
\begin{center}
\begin{longtable}{l c c >{\centering\arraybackslash}m{5cm} >{\centering\arraybackslash}m{5cm}}
\caption{A recipe for going from effective operators to models. For each operator closable in two loops or less, we list the key details required for model building. We list the inferred upper bound on the scale of new physics $\Lambda_{\nu}$ from Ref.~\cite{dgj}, followed by the technique required to close off the loops: a, b or c as described in Sec.~\ref{sec:loop} and displayed in example form in Figs.~\ref{fig:O49}, \ref{fig:O11b} and \ref{fig:O8}, respectively. Finally we list the 1- and 2-loop topologies available for the UV completion, with reference to the figures from the above analysis.} \\
\centering ${\cal O}$ & $\Lambda_{\nu}$ (TeV) & Loop Closure & 1-loop Topologies & 2-loop Topologies \\ \hline \hline
\label{tab:summary}
\endfirsthead

\centering ${\cal O}$ & $\Lambda_{\nu}$ (TeV) & Loop Closure & 1-loop Topologies & 2-loop Topologies \\ \hline \hline
\endhead

\multicolumn{5}{c}{Four-fermion Operators} \\ \hline
2 & 4$\times$10$^{\textrm{7}}$ & b  & Fig.~\ref{fig:scalaropen} & Fig.~\ref{fig:ExternalLoop} \\
3a & 2$\times$10$^{\textrm{5}}$ & c & - & Fig.~\ref{fig:O8UV}  \\
3b & 1$\times$10$^{\textrm{8}}$ & b & Fig.~\ref{fig:scalaropen} & Fig.~\ref{fig:ExternalLoop} \\
4a & 4$\times$10$^{\textrm{9}}$ & b & Fig.~\ref{fig:O6}  & Fig.~\ref{fig:ExternalLoop}  \\ 
4b & 6$\times$10$^{\textrm{6}}$ & c & - &  Fig.~\ref{fig:O8UV} \\ 
5 & 6$\times$10$^{\textrm{5}}$ & b & Fig.~\ref{fig:scalaropen} & Fig.~\ref{fig:ExternalLoop} \\
6 & 2$\times$10$^{\textrm{7}}$ & b & Fig.~\ref{fig:O6} & Fig.~\ref{fig:ExternalLoop} \\ 
7 & 4$\times$10$^{\textrm{2}}$ & c & - & Fig.~\ref{fig:O8UV}  \\ 
8 & 6$\times$10$^{\textrm{3}}$ & c & - & Fig.~\ref{fig:O8UV}  \\ 
61 & 2$\times$10$^{\textrm{5}}$ & b & Fig.~\ref{fig:scalaropen} & Fig.~\ref{fig:ExternalLoop} \\
66 & 6$\times$10$^{\textrm{5}}$ & b & Fig.~\ref{fig:scalaropen} & Fig.~\ref{fig:ExternalLoop} \\
71 & 2$\times$10$^{\textrm{7}}$ & b & Fig.~\ref{fig:scalaropen} & Fig.~\ref{fig:ExternalLoop} \\ \hline 
\multicolumn{5}{c}{9D Six-fermion Operators} \\ \hline
9 & 3$\times$10$^{\textrm{3}}$ & b & - & Fig.~\ref{fig:6fscalar} and \ref{fig:6ffermionclosure} B-D \\ 
10 & 6$\times$10$^{\textrm{3}}$ & b & - & Fig.~\ref{fig:6fscalar} and \ref{fig:6ffermionclosure} B-D \\ 
11b & 2$\times$10$^{\textrm{4}}$ & b & - & Fig.~\ref{fig:6fscalar} and \ref{fig:6ffermionclosure} B-D \\ 
12a & 2$\times$10$^{\textrm{7}}$ & b & - & Fig.~\ref{fig:6fscalar} and \ref{fig:6ffermionclosure} B-D \\ 
13 & 2$\times$10$^{\textrm{5}}$ & b & - & Fig.~\ref{fig:6fscalar} and \ref{fig:6ffermionclosure} B-D \\ 
14b & 6$\times$10$^{\textrm{5}}$ & b & - & Fig.~\ref{fig:6fscalar} and \ref{fig:6ffermionclosure} B-D \\ \hline 
\multicolumn{5}{c}{11D Six-fermion Operators} \\ \hline
21a-b & 2$\times$10$^{\textrm{3}}$ & b & - & Fig.~\ref{fig:6fscalar} and \ref{fig:6ffermionclosure} \\
22 & 6$\times$10$^{\textrm{6}}$ & a & - & Fig.~\ref{fig:6fscalar} and \ref{fig:6ffermionclosure} \\
23 & 40 & b & - & Fig.~\ref{fig:6fscalar} and \ref{fig:6ffermionclosure} \\
25 & 4$\times$10$^{\textrm{3}}$ & b & - & Fig.~\ref{fig:6fscalar} and \ref{fig:6ffermionclosure} \\
26b & 40 & b & - & Fig.~\ref{fig:6fscalar} and \ref{fig:6ffermionclosure} \\
27a-b & 6$\times$10$^{\textrm{6}}$ & a & - & Fig.~\ref{fig:6fscalar} and \ref{fig:6ffermionclosure} \\
29a & 2$\times$10$^{\textrm{5}}$ & b & - & Fig.~\ref{fig:6fscalar} and \ref{fig:6ffermionclosure} \\
30b & 2$\times$10$^{\textrm{3}}$ & b & - & Fig.~\ref{fig:6fscalar} and \ref{fig:6ffermionclosure} \\
31a-b & 4$\times$10$^{\textrm{3}}$ & b & - & Fig.~\ref{fig:6fscalar} and \ref{fig:6ffermionclosure} \\
33 & 6$\times$10$^{\textrm{6}}$ & a & - & Fig.~\ref{fig:6fscalar} and \ref{fig:6ffermionclosure} \\
39a-d & 6$\times$10$^{\textrm{6}}$ & a & - & Fig.~\ref{fig:6fscalar} and \ref{fig:6ffermionclosure} \\
40a-j & 6$\times$10$^{\textrm{6}}$ & a & - & Fig.~\ref{fig:6fscalar} and \ref{fig:6ffermionclosure} \\
41a-b & 6$\times$10$^{\textrm{6}}$ & a & - & Fig.~\ref{fig:6fscalar} and \ref{fig:6ffermionclosure} \\
42a-b & 6$\times$10$^{\textrm{6}}$ & a & - & Fig.~\ref{fig:6fscalar} and \ref{fig:6ffermionclosure} \\
44a-b & 6$\times$10$^{\textrm{6}}$ & a & - & Fig.~\ref{fig:6fscalar} and \ref{fig:6ffermionclosure} \\
44d & 6$\times$10$^{\textrm{6}}$ & a & - & Fig.~\ref{fig:6fscalar} and \ref{fig:6ffermionclosure} \\
45 & 6$\times$10$^{\textrm{6}}$ & a & - & Fig.~\ref{fig:6fscalar} and \ref{fig:6ffermionclosure} \\
46 & 6$\times$10$^{\textrm{6}}$ & a & - & Fig.~\ref{fig:6fscalar} and \ref{fig:6ffermionclosure} \\
47a-e & 6$\times$10$^{\textrm{6}}$ & a & - & Fig.~\ref{fig:6fscalar} and \ref{fig:6ffermionclosure} \\
47g-j & 6$\times$10$^{\textrm{6}}$ & a & - & Fig.~\ref{fig:6fscalar} and \ref{fig:6ffermionclosure} \\
48 & 6$\times$10$^{\textrm{6}}$ & a & - & Fig.~\ref{fig:6fscalar} and \ref{fig:6ffermionclosure} \\
49 & 6$\times$10$^{\textrm{6}}$ & a & - & Fig.~\ref{fig:6fscalar} and \ref{fig:6ffermionclosure} \\
51 & 6$\times$10$^{\textrm{6}}$ & a & - & Fig.~\ref{fig:6fscalar} and \ref{fig:6ffermionclosure} \\
62 & 20 & b & - & Fig.~\ref{fig:6fscalar} and \ref{fig:6ffermionclosure} \\
63b & 40 & b & - & Fig.~\ref{fig:6fscalar} and \ref{fig:6ffermionclosure} \\
64a & 2$\times$10$^{\textrm{3}}$ & b & - & Fig.~\ref{fig:6fscalar} and \ref{fig:6ffermionclosure} \\
67 & 40 & b & - & Fig.~\ref{fig:6fscalar} and \ref{fig:6ffermionclosure} \\
68b & 1$\times$10$^{\textrm{2}}$ & b & - & Fig.~\ref{fig:6fscalar} and \ref{fig:6ffermionclosure} \\
69a & 4$\times$10$^{\textrm{3}}$ & b & - & Fig.~\ref{fig:6fscalar} and \ref{fig:6ffermionclosure} \\
72 & 2$\times$10$^{\textrm{3}}$ & b & - & Fig.~\ref{fig:6fscalar} and \ref{fig:6ffermionclosure} \\
73b & 4$\times$10$^{\textrm{3}}$ & b & - & Fig.~\ref{fig:6fscalar} and \ref{fig:6ffermionclosure} \\
74a & 2$\times$10$^{\textrm{5}}$ & b & - & Fig.~\ref{fig:6fscalar} and \ref{fig:6ffermionclosure} \\ \hline
 \end{longtable}
\end{center}

\section{Additional discussion and Conclusions}
\label{sec:conc}

We have outlined how to construct all minimal neutrino mass diagrams from four and six fermi operators in the list of Babu and Leung \cite{babuleung}.  After choosing an operator, one simply has to close the loops as in Sec.~\ref{sec:loop} and UV complete the vertex as outlined in Secs.~\ref{sec:4f} and \ref{sec:6f}, and all these results have been collected in Sec.~\ref{sec:summary}.  It is hoped this addition to the growing literature on a systematic bottom-up approach to the problem of neutrino mass will help provide a clearer path through the allowable model space.\footnote{A precise statement of the scope of our analysis is presented at the end of Sec.~\ref{sec:opslist}.}  In addition, the combination of our recipe for constructing neutrino mass diagrams and the work of Ref.~\cite{dgj} on the testable scale of various operators, should allow for the construction of models with interesting LHC phenomenology.

Our analysis reveals that 11D operators in general give rise to the largest number of graphs, which naively suggests these operators might be associated with a substantial model space, which is so far unexplored. This is an interesting space given that if one were able to rule out 11D operators as the origin of neutrino mass, the list of effective operators would be reduced from 75 to 23, of which only 17 can be closed in 2-loops or less. In such an event it may actually be tractable to write down every possible minimal neutrino mass model and test them individually.

\begin{figure}
\centering
\includegraphics[height=0.40\columnwidth]{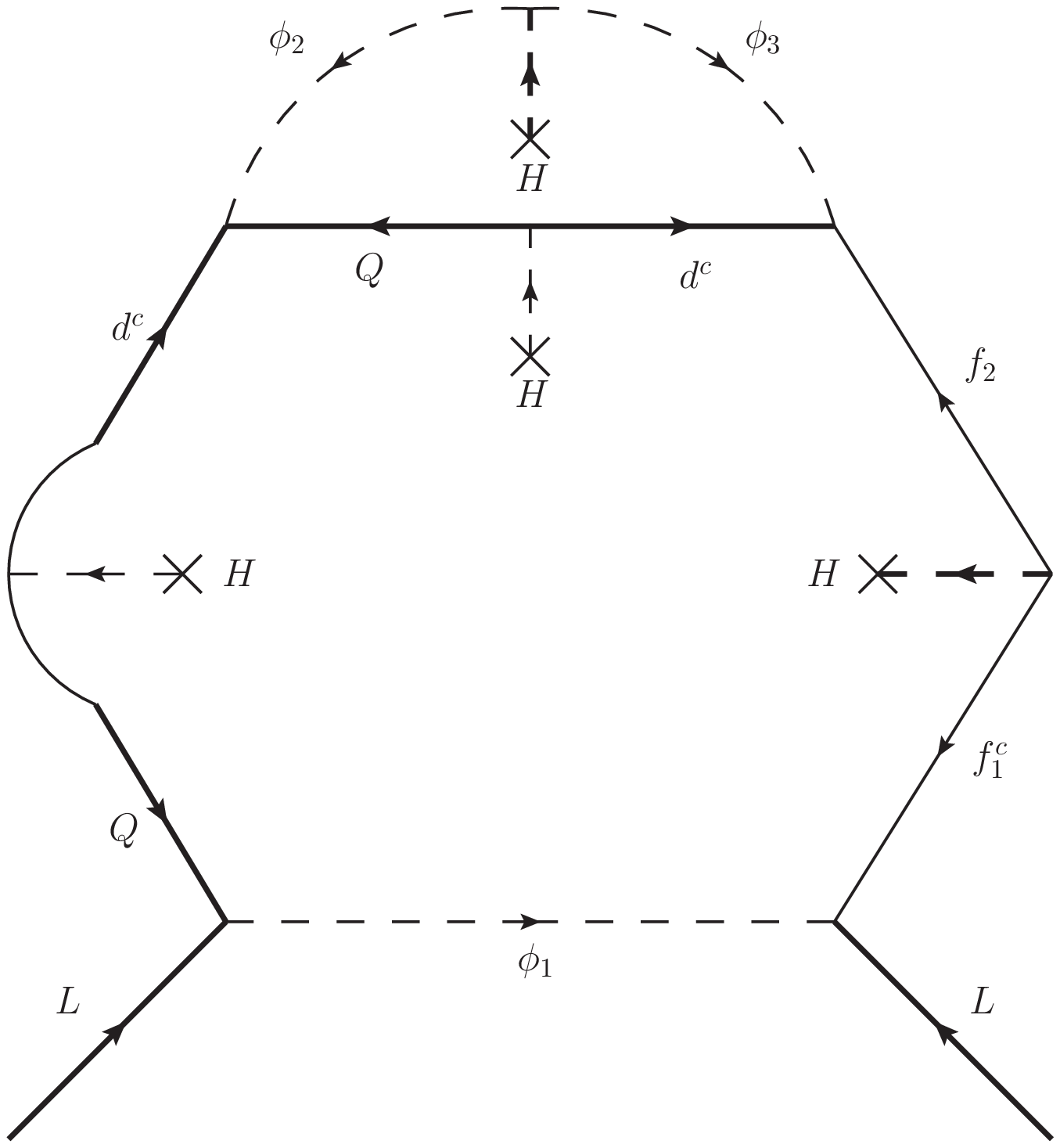}
\includegraphics[height=0.40\columnwidth]{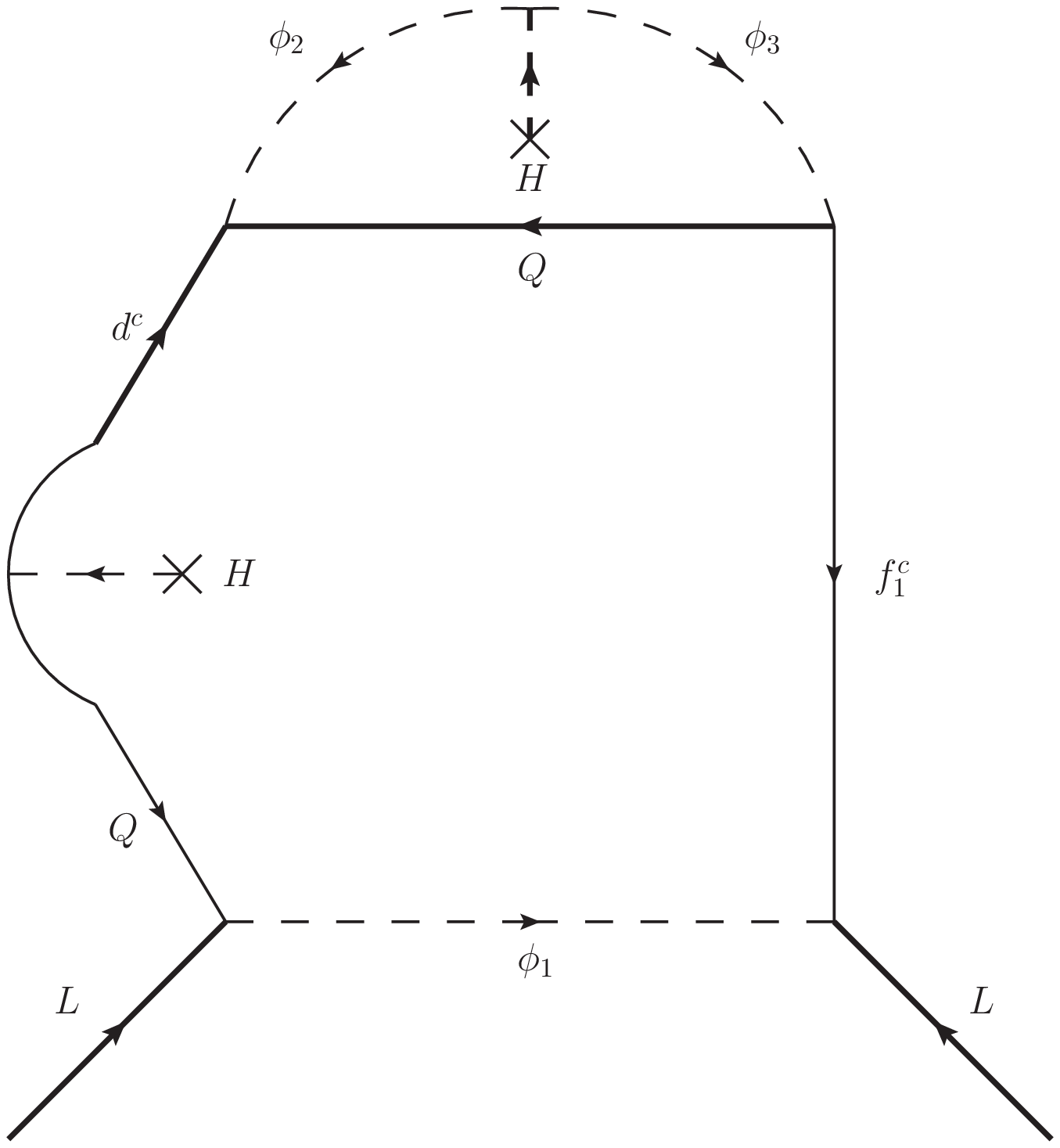}
\caption{Diagram generated from ${\cal O}_{68b}$ using diagram B (left) and a second 2-loop diagram the full model allows, which can be integrated back to ${\cal O}_3$ (right).}
\label{fig:O68b}
\end{figure}

In general it appears to be difficult to write down a complete model that originates purely from an 11D operator.  For example consider ${\cal O}_{68b} = L^i L^j Q^k d^c H^l Q^r d^c \overline{H}_r \epsilon_{ik} \epsilon_{jl}$.  On the left of Fig.~\ref{fig:O68b} we show a graph derived from this operator using diagram B from Fig.~\ref{fig:6ffermionclosure}.  From this graph one can derive the transformation properties of the five new fields and then write down the most general Lagrangian allowed by gauge invariance; any ambiguities in the quantum numbers of the new fields are resolved so as to minimise the number of new terms in the Lagrangian.  In this specific example it turns out the Lagrangian allows a second diagram that generates neutrino mass, which we have depicted on the right of Fig.~\ref{fig:O68b}.  The second graph can be integrated back to the 7D operator ${\cal O}_3$ (to see this note that we treat $\overline{Q} d H$ as a massive down type quark propagator when evaluating the amplitude, so this can be integrated back to $d^c$).  In fact one can calculate that the ${\cal O}_3$ diagram will dominate the induced neutrino mass over essentially the entire parameter space of the model, making the 11D aspects of this model negligible.

One might suspect that the problem with the above is that ${\cal O}_{68b}$ is a product operator, specifically ${\cal O}_{68b} = \left( {\cal O}_{3b} \right) \left( Q^r d^c \overline{H}_r \right)$ and this problem has arisen as we have not used non-trivial Lorentz contractions to prevent inducing ${\cal O}_3$, as suggested in Ref.~\cite{babuleung}.  In fact non-trivial Lorentz contractions are not possible for this operator, however the objection remains.  Nevertheless we found this process repeated itself for several other 11D operators, including those that were not product operators.  For example a model constructed from ${\cal O}_{31a}$ using diagram C from Fig.~\ref{fig:6ffermionclosure} induced graphs that integrated back to ${\cal O}_4$.  The problem may be that these operators are often similar to lower dimensional counterparts, but with additional structure.  In such situations at least some of the new particles introduced in the UV completion will have appeared in graphs from lower dimensional operators, and it appears these are often enough to generate the diagrams associated with them.  Of course this hardly amounts to a proof that 11D operators are ignorable from the perspective of neutrino mass, which remains an open and interesting question.

\begin{acknowledgments}

We thank K. S. Babu and A. de Gouv\^{e}a for many very useful discussions, K. S. Babu for detailed comments on an earlier draft, and also Y. Cai and M. A. Schmidt for feedback on that draft.  We also thank J. M. No for enlightening communications about effective operators containing gauge fields.  This work was supported in part by the Australian Research Council.

\end{acknowledgments}

\appendix*

\section{List of effective operators}

The list of effective operators up to mass dimension 11 is reproduced here for the convenience of the reader.  All of the fermi fields are left-handed, with $L$ and $Q$ being the lepton and quark doublets, respectively, and $e^c$, $u^c$ and $d^c$ being the isosinglet charged antilepton, up antiquark and down antiquark, respectively. The scalar $H$ is the Higgs doublet, with the convention that its hypercharge is opposite that of $L$; $\overline{H}$ is then the conjugate.  Lower case letters from the middle of the Roman alphabet are weak isospin indices.  Colour indices are not indicated.  The compact notation leaves the Lorentz structure to be inferred.  Thus
\begin{equation}
{\cal O}_2 \equiv L L L e^c H = \overline{(L_L)^c} L_L \overline{(L_L)^c} (e_R)^c H = \overline{(L_L)^c} L_L \overline{e}_R L_L H,
\end{equation}
and so on.  An overbar on a fermi field when the compact notation is being used means a right-handed field, for example $\overline{Q} = (Q_L)^c$.  Thus
\begin{equation}
{\cal O}_4 \equiv LL \overline{Q} \overline{u}^c H = \overline{(L_L)^c} L_L \overline{((Q_L)^c)^c} ((u_R)^c)^c H = \overline{(L_L)^c} L_L \overline{Q}_L u_R H.
\end{equation}
Note that vector, axial-vector and tensor Lorentz contractions are not relevant for the present analysis.

%\newpage

Here is the list:
\begin{align}
{\cal O}_2&= L^i L^j L^k e^c H^l \epsilon_{ij} \epsilon_{kl}
\nonumber \\
{\cal O}_3&= \{L^i L^j Q^k d^c H^l \epsilon_{ij} \epsilon_{kl},
~~L^i L^j Q^k d^c H^l \epsilon_{ik} \epsilon_{jl}\}  \nonumber \\
{\cal O}_4&= \{L^i L^j \overline{Q}_i \overline{u}^c H^k \epsilon_{jk},~~
L^i L^j \overline{Q}_k\overline{u}^cH^k \epsilon_{ij}\}
\nonumber \\
{\cal O}_5 &= L^i L^j Q^k d^c H^l H^m \overline{H}_i \epsilon_{jl}
\epsilon_{km} \nonumber \\
{\cal O}_6 &= L^i L^j \overline{Q}_k\overline{u}^cH^l H^k \overline{H}_i
\epsilon_{jl} \nonumber \\
{\cal O}_{7} &= L^iQ^j \overline{e}^c\overline{Q}_kH^k H^l H^m 
\epsilon_{il} \epsilon_{jm} \nonumber \\
{\cal O}_{8} &= L^i \overline{e}^c \overline{u}^c d^c H^j \epsilon_{ij}
\nonumber \\
{\cal O}_{9}&=L^i L^j L^k e^c L^l e^c \epsilon_{ij}
\epsilon_{kl} \nonumber \\
{\cal O}_{10}&=L^i L^j L^k e^c Q^l d^c \epsilon_{ij}
\epsilon_{kl} \nonumber \\
{\cal O}_{11}&=\{L^i L^j Q^k d^c Q^l d^c \epsilon_{ij}
\epsilon_{kl},~~L^i L^j Q^k d^c Q^l d^c \epsilon_{ik}
\epsilon_{jl}\} \nonumber \\
{\cal O}_{12}&=\{L^i L^j \overline{Q}_i \overline{u}^c \overline{Q_j}
\overline{u}^c,~~L^i L^j \overline{Q}_k \overline{u}^c \overline{Q}_l \overline{u}^c
\epsilon_{ij} \epsilon^{kl} \}
\nonumber \\
{\cal O}_{13}&=L^i L^j \overline{Q}_i \overline{u}^cL^l e^c
\epsilon_{jl} \nonumber \\
{\cal O}_{14}&=\{L^i L^j \overline{Q}_k \overline{u}^c Q^k d^c
\epsilon_{ij},~~L^i L^j \overline{Q}_i \overline{u}^c Q^l d^c
\epsilon_{jl}\} \nonumber \\
{\cal O}_{15} &= L^i L^j L^k d^c \overline{L}_i \overline{u}^c
\epsilon_{jk} \nonumber \\
{\cal O}_{16} &= L^i L^j e^c d^c \overline{e}^c \overline{u}^c
\epsilon_{ij} \nonumber \\
{\cal O}_{17}&= L^i L^j d^c d^c \overline{d}^c \overline{u}^c
\epsilon_{ij} \nonumber \\
{\cal O}_{18}&= L^i L^j d^c u^c \overline{u}^c \overline{u}^c
\epsilon_{ij} \nonumber \\
{\cal O}_{19}&= L^i Q^j d^c d^c \overline{e}^c \overline{u}^c
\epsilon_{ij} \nonumber \\
{\cal O}_{20} &= L^i d^c \overline{Q}_i \overline{u}^c \overline{e}^c
\overline{u}^c \nonumber \\
{\cal O}_{23} &= L^i L^jL^k e^c \overline{Q}_k\overline{d}^cH^l
H^m \epsilon_{il} \epsilon_{jm} \nonumber \\
{\cal O}_{24} &= \{L^i L^j Q^k d^c Q^l d^c H^m \overline{H}_i
\epsilon_{jk} \epsilon_{lm},~~~ L^i L^j Q^k d^c Q^l d^c
H^m \overline{H}_i \epsilon_{jm} \epsilon_{kl}\} \nonumber \\
{\cal O}_{25} &=  
L^i L^j Q^k d^c Q^l u^c H^m H^n \epsilon_{im}
\epsilon_{jn} \epsilon_{kl} \nonumber \\
{\cal O}_{26} &= \{L^i L^j Q^k d^c \overline{L}_i \overline{e}^c
H^l H^m \epsilon_{jl} \epsilon_{km},~~~L^i L^j Q^k d^c
\overline{L}_k \overline{e}^c H^l H^m \epsilon_{il} \epsilon_{jm}\}
\nonumber \\
{\cal O}_{27} &= \{L^i L^j Q^k d^c \overline{Q}_i\overline{d}^c H^l
H^m \epsilon_{jl} \epsilon_{km},~~~L^i L^j Q^k d^c
\overline{Q}_k\overline{d}^c H^l H^m \epsilon_{il} \epsilon_{jm}\}
\nonumber \\
{\cal O}_{28} &= \{L^i L^j Q^k d^c \overline{Q}_j \overline{u}^c
H^l \overline{H}_i \epsilon_{kl},~~~L^i L^j Q^k d^c \overline{Q}_k
\overline{u}^c H^l \overline{H}_i \epsilon_{jl}, \nonumber \\
&~ L^i L^j Q^k d^c \overline{Q}_l \overline{u}^c H^l \overline{H}_i
\epsilon_{jk}\} \nonumber \\
{\cal O}_{29} &= \{L^i L^j Q^k u^c \overline{Q}_k \overline{u}^c 
H^l H^m \epsilon_{il} \epsilon_{jm},~~~L^i L^j Q^k u^c
\overline{Q}_l \overline{u}^c H^l H^m \epsilon_{ik} \epsilon_{jm}\}
\nonumber \\
{\cal O}_{30} &= \{L^i L^j \overline{L}_i \overline{e}^c
\overline{Q}_k\overline{u}^c H^k H^l \epsilon_{jl},~~~L^i L^j
\overline{L}_m \overline{e}^c \overline{Q}_n \overline{u}^c H^k H^l
\epsilon_{ik} \epsilon_{jl} \epsilon^{mn}\} \nonumber \\
{\cal O}_{31} &= \{L^i L^j \overline{Q}_i\overline{d}^c
\overline{Q}_k\overline{u}^c H^k H^l \epsilon_{jl},~~~L^i L^j
\overline{Q}_m\overline{d}^c \overline{Q}_n\overline{u}^c H^k H^l
\epsilon_{ik} \epsilon_{jl} \epsilon^{mn}\} \nonumber \\
{\cal O}_{32} &= \{L^i L^j \overline{Q}_j \overline{u}^c
\overline{Q}_k \overline{u}^c H^k \overline{H}_i,~~~L^i L^j
\overline{Q}_m \overline{u}^c \overline{Q}_n \overline{u}^c H^k
\overline{H}_i \epsilon_{jk} \epsilon^{mn}\} \nonumber \\
{\cal O}_{33} &= \overline{e}^c \overline{e}^c L^i L^j e^c e^c H^k
H^l \epsilon_{ik} \epsilon_{jl} \nonumber \\
{\cal O}_{34} &= \overline{e}^c \overline{e}^c L^i Q^j e^c d^c H^k
H^l \epsilon_{ik} \epsilon_{jl} \nonumber \\
{\cal O}_{35} &= \overline{e}^c \overline{e}^c L^i e^c \overline{Q}_j
\overline{u}^c H^j H^k \epsilon_{ik} \nonumber \\
{\cal O}_{36} &= \overline{e}^c \overline{e}^c Q^i d^c Q^j d^c H^k
H^l \epsilon_{ik} \epsilon_{jl} \nonumber \\
{\cal O}_{37} &= \overline{e}^c \overline{e}^c Q^i d^c \overline{Q}_j
\overline{u}^c H^j H^k \epsilon_{ik} \nonumber \\
{\cal O}_{38} &= \overline{e}^c \overline{e}^c \overline{Q}_i \overline{u}^c
\overline{Q}_j \overline{u}^c H^i H^j \nonumber \\
{\cal O}_{39} &= \{L^i L^j L^k L^l \overline{L}_i \overline{L}_j
H^m H^n \epsilon_{jm} \epsilon_{kl},~~~L^i L^j L^k L^l
\overline{L}_m \overline{L}_n H^m H^n \epsilon_{ij} \epsilon_{kl},
\nonumber \\
&~ L^i L^j L^k L^l \overline{L}_i \overline{L}_m H^m H^n
\epsilon_{jk} \epsilon_{ln},~~~~L^i L^j L^k L^l \overline{L}_p
\overline{L}_q H^m H^n \epsilon_{ij} \epsilon_{km} \epsilon_{ln}
\epsilon^{pq}\} \nonumber \\
{\cal O}_{40} &= \{L^i L^j L^k Q^l \overline{L}_i \overline{Q}_j
H^m H^n \epsilon_{km} \epsilon_{ln},~~~L^i L^j L^k Q^l
\overline{L}_i \overline{Q}_l H^m H^n \epsilon_{jm} \epsilon_{kn},
\nonumber \\
&~ L^i L^j L^k Q^l \overline{L}_l \overline{Q}_i H^m H^n
\epsilon_{jm} \epsilon_{kn}, ~~~~L^i L^j L^k Q^l \overline{L}_i
\overline{Q}_m H^m H^n \epsilon_{jk} \epsilon_{ln}, \nonumber \\
&~ L^i L^j L^k Q^l \overline{L}_i \overline{Q}_m H^m H^n \epsilon_{jl}
\epsilon_{kn}, ~~~L^i L^j L^k Q^l \overline{L}_m \overline{Q}_i H^m
H^n \epsilon_{jk} \epsilon_{ln}, \nonumber \\
&~ L^i L^j L^k Q^l \overline{L}_m \overline{Q}_i H^m H^n \epsilon_{jl}
\epsilon_{kn},~~~~L^i L^j L^k Q^l \overline{L}_m \overline{Q}_n H^m H^n
\epsilon_{ij} \epsilon_{kl}, \nonumber \\
&~ L^i L^j L^k Q^l \overline{L}_m \overline{Q}_n H^p H^q \epsilon_{ip}
\epsilon_{jq} \epsilon_{kl} \epsilon^{mn},~~~
L^i L^j L^k Q^l \overline{L}_m \overline{Q}_n H^p H^q \epsilon_{ip}
\epsilon_{lq} \epsilon_{jk} \epsilon^{mn}\} \nonumber \\
{\cal O}_{41} &= \{L^i L^j L^k d^c \overline{L}_i \overline{d}^c H^l
H^m \epsilon_{jl} \epsilon_{km},~~~L^i L^j L^k d^c \overline{L}_l
\overline{d}^c H^l H^m \epsilon_{ij} \epsilon_{km}\} \nonumber \\
{\cal O}_{42} &= \{L^i L^j L^k u^c \overline{L}_i \overline{u}^c H^l
H^m \epsilon_{jl} \epsilon_{km},~~~L^i L^j L^k u^c \overline{L}_l
\overline{u}^c H^l H^m \epsilon_{ij} \epsilon_{km}\} \nonumber \\
{\cal O}_{43} &= \{L^i L^j L^k d^c \overline{L}_l\overline{u}^c H^l
\overline{H}_i \epsilon_{jk},~~~L^i L^j L^k d^c \overline{L}_j\overline{u}^c
H^l \overline{H}_i \epsilon_{kl}, \nonumber \\
&~ L^i L^j L^k d^c \overline{L}_l\overline{u}^c H^m \overline{H}_n
\epsilon_{ij} \epsilon_{km} \epsilon^{ln}\} \nonumber \\
{\cal O}_{44} &= \{L^i L^j Q^k e^c \overline{Q}_i \overline{e}^c H^l
H^m \epsilon_{jl} \epsilon_{km},~~~L^i L^j Q^k e^c \overline{Q}_k
\overline{e}^c H^l H^m \epsilon_{il} \epsilon_{jm}, \nonumber \\
&~ L^i L^j Q^k e^c \overline{Q}_l \overline{e}^c H^l H^m \epsilon_{ij}
\epsilon_{km},~~~L^i L^j Q^k e^c \overline{Q}_l \overline{e}^c H^l H^m
\epsilon_{ik} \epsilon_{jm}\} \nonumber \\
{\cal O}_{45} &= L^i L^j e^c d^c \overline{e}^c \overline{d}^c H^k H^l
\epsilon_{ik} \epsilon_{jl} \nonumber \\
{\cal O}_{46} &= L^i L^j e^c u^c \overline{e}^c \overline{u}^c H^k H^l
\epsilon_{ik} \epsilon_{jl} \nonumber \\
{\cal O}_{47} &= \{L^i L^j Q^k Q^l \overline{Q}_i\overline{Q}_j H^m H^n
\epsilon_{km} \epsilon_{ln},~~~L^i L^j Q^k Q^l \overline{Q}_i
\overline{Q}_k H^m H^n \epsilon_{jm} \epsilon_{ln}, \nonumber \\
&~ L^i L^j Q^k Q^l \overline{Q}_k\overline{Q}_l H^m H^n \epsilon_{im}
\epsilon_{jn},~~~L^i L^j Q^k Q^l \overline{Q}_i\overline{Q}_m H^m H^n
\epsilon_{jk} \epsilon_{ln}, \nonumber \\
&~ L^i L^j Q^k Q^l \overline{Q}_i\overline{Q}_m H^m H^n \epsilon_{jn}
\epsilon_{kl},~~~L^i L^j Q^k Q^l \overline{Q}_k\overline{Q}_m H^m H^n
\epsilon_{ij} \epsilon_{ln}, \nonumber \\
&~ L^i L^j Q^k Q^l \overline{Q}_k\overline{Q}_m H^m H^n \epsilon_{il}
\epsilon_{jn},~~~L^i L^j Q^k Q^l \overline{Q}_p\overline{Q}_q H^m H^n
\epsilon_{ij} \epsilon_{km} \epsilon_{ln} \epsilon^{pq}
\nonumber \\
&~ L^i L^j Q^k Q^l \overline{Q}_p\overline{Q}_q H^m H^n \epsilon_{ik}
\epsilon_{jm} \epsilon_{ln} \epsilon^{pq},~~~
L^i L^j Q^k Q^l \overline{Q}_p\overline{Q}_q H^m H^n \epsilon_{im}
\epsilon_{jn} \epsilon_{kl} \epsilon^{pq}\} \nonumber \\
{\cal O}_{48} &= L^i L^j d^c d^c \overline{d}^c \overline{d}^c H^k
H^l \epsilon_{ik} \epsilon_{jl} \nonumber \\
{\cal O}_{49} &= L^i L^j d^c u^c \overline{d}^c \overline{u}^c H^k
H^l \epsilon_{ik} \epsilon_{jl} \nonumber \\
{\cal O}_{50} &= L^i L^j d^c d^c \overline{d}^c \overline{u}^c H^k
\overline{H}_i \epsilon_{jk} \nonumber \\
{\cal O}_{51} &= L^i L^j u^c u^c \overline{u}^c \overline{u}^c H^k
H^l \epsilon_{ik} \epsilon_{jl} \nonumber \\
{\cal O}_{52} &= L^i L^j d^c u^c \overline{u}^c \overline{u}^c H^k
\overline{H}_i \epsilon_{jk} \nonumber \\
{\cal O}_{53} &= L^i L^j d^c d^c \overline{u}^c \overline{u}^c
\overline{H}_i \overline{H}_j \nonumber \\
{\cal O}_{54} &= \{L^i Q^j Q^k d^c \overline{Q}_i \overline{e}^c
H^l H^m \epsilon_{jl} \epsilon_{km},~~~L^i Q^j Q^k d^c
\overline{Q}_j \overline{e}^c H^l H^m \epsilon_{il} \epsilon_{km},
\nonumber \\
&~ L^i Q^j Q^k d^c \overline{Q}_l \overline{e}^c H^l H^m \epsilon_{im}
\epsilon_{jk},~~~L^i Q^j Q^k d^c \overline{Q}_l \overline{e}^c H^l
H^m \epsilon_{ij} \epsilon_{km}\} \nonumber \\
{\cal O}_{55} &= \{L^i Q^j \overline{Q}_i \overline{Q}_k \overline{e}^c
\overline{u}^c H^k H^l \epsilon_{jl},~~~L^i Q^j \overline{Q}_j
\overline{Q}_k \overline{e}^c \overline{u}^c H^k H^l \epsilon_{il},
\nonumber \\
&~L^i Q^j \overline{Q}_m \overline{Q}_n \overline{e}^c \overline{u}^c H^k H^l
\epsilon_{ik} \epsilon_{jl} \epsilon^{mn}\} \nonumber \\
{\cal O}_{56} &= L^i Q^j d^c d^c \overline{e}^c \overline{d}^c H^k
H^l \epsilon_{ik} \epsilon_{jl} \nonumber \\
{\cal O}_{57} &= L^i d^c \overline{Q}_j \overline{u}^c \overline{e}^c
\overline{d}^c H^j H^k \epsilon_{ik} \nonumber \\
{\cal O}_{58} &= L^i u^c \overline{Q}_j \overline{u}^c \overline{e}^c
\overline{u}^c H^j H^k \epsilon_{ik} \nonumber \\
{\cal O}_{59} &= L^i Q^j d^c d^c \overline{e}^c \overline{u}^c
H^k \overline{H}_i \epsilon_{jk} \nonumber \\
{\cal O}_{60} &= L^i d^c \overline{Q}_j \overline{u}^c \overline{e}^c
\overline{u}^c H^j \overline{H}_i \nonumber \\
{\cal O}_{61} &= L^i L^j H^k H^l L^r e^c \overline{H}_r \epsilon_{ik} 
\epsilon_{jl} \nonumber \\
{\cal O}_{62} &= L^i L^j L^k e^c H^l L^r e^c \overline{H}_r \epsilon_{ij} 
\epsilon_{kl} \nonumber \\
{\cal O}_{63} &=\{L^i L^j Q^k d^c H^l L^r e^c \overline{H}_r \epsilon_{ij}
\epsilon{kl},~~L^i L^j Q^k d^c H^l L^r e^c \overline{H}_r \epsilon_{ik}
\epsilon{jl} \} \nonumber \\
{\cal O}_{64} &=\{L^i L^j \overline{Q}_i \overline{u}^c H^k L^r e^c \overline{H}_r \epsilon_{jk},~~L^i L^j \overline{Q}_k \overline{u}^c H^k L^r e^c \overline{H}_r 
\epsilon_{ij} \} \nonumber \\
{\cal O}_{65} &= L^i \overline{e}^c \overline{u}^c d^c H^j L^r e^c \overline{H}_r 
\epsilon_{ij} \nonumber \\
{\cal O}_{66} &= L^i L^j H^k H^l \epsilon_{ik} Q^r d^c \overline{H}_r \epsilon_{jl} 
\nonumber \\
{\cal O}_{67} &= L^i L^j L^k e^c H^l Q^r d^c \overline{H}_r \epsilon_{ij} 
\epsilon_{kl} \nonumber \\
{\cal O}_{68} &=\{L^i L^j Q^k d^c H^l Q^r d^c \overline{H}_r \epsilon_{ij} 
\epsilon_{kl},~~L^i L^j Q^k d^c H^l Q^r d^c \overline{H}_r \epsilon_{ik} 
\epsilon_{jl} \} \nonumber \\
{\cal O}_{69} &=\{L^i L^j \overline{Q}_i \overline{u}^c H^k Q^r d^c \overline{H}_r 
\epsilon_{jk},~~L^i L^j \overline{Q}_k \overline{u}^c H^k Q^r d^c \overline{H}_r 
\epsilon_{ij} \} \nonumber \\
{\cal O}_{70} &= L^i \overline{e}^c \overline{u}^c d^c H^j Q^r d^c \overline{H}_r
\epsilon_{ij} \nonumber \\
{\cal O}_{71} &= L^i L^j H^k H^l Q^r u^c H^s \epsilon_{rs} \epsilon_{ij} 
\epsilon_{jl} \nonumber \\
{\cal O}_{72} &= L^i L^j L^k e^c H^l Q^r u^c H^s \epsilon_{rs} \epsilon_{ij}
\epsilon_{kl} \nonumber \\
{\cal O}_{73} &= \{L^i L^j Q^k d^c H^l Q^r u^c H^s \epsilon_{rs} \epsilon_{ij}
\epsilon_{kl},~~L^i L^j Q^k d^c H^l Q^r u^c H^s \epsilon_{rs} \epsilon_{ik}
\epsilon_{jl} \} \nonumber \\
{\cal O}_{74} &= \{L^i L^j \overline{Q}_i \overline{u}^c H^k Q^r u^c H^s 
\epsilon_{rs} \epsilon_{jk},~~L^i L^j \overline{Q}_k \overline{u}^c H^k Q^r u^c 
H^s \epsilon_{rs} \epsilon_{ij} \} \nonumber \\
{\cal O}_{75} &= L^i \overline{e}^c \overline{u}^c d^c H^j Q^r u^c H^s 
\epsilon_{rs} \epsilon_{ij} \nonumber
%\label{O6fd11}
\end{align}

\begin{spacing}{0.5}
\bibliographystyle{ieeetr}
\bibliography{bib}
\end{spacing}

\end{document}